\documentclass[twocolumn,times,tighten]{aastex63}

\usepackage{amsmath, amssymb}
\usepackage{xspace}
\usepackage{datetime}

\usepackage{cleveref}

\usepackage[normalem]{ulem}

\newcommand{\kms}{\ensuremath{\,\text{km}~\text{s}^{-1}}}

\newcommand{\oiii}{\ensuremath{\text{[\ion{O}{3}]}}\xspace}
\newcommand{\oiiitot}{\ensuremath{\text{[\ion{O}{3}]}\lambda\lambda4959,5007}\xspace}
\newcommand{\oiiione}{\ensuremath{\text{[\ion{O}{3}]}\lambda4959}\xspace}
\newcommand{\oiiitwo}{\ensuremath{\text{[\ion{O}{3}]}\lambda5007}\xspace}
\newcommand{\oiiidirect}{\ensuremath{\text{[\ion{O}{3}]}\lambda4363}\xspace}
\newcommand{\oii}{\ensuremath{\text{[\ion{O}{2}]}\lambda3727}\xspace}
\newcommand{\oiitot}{\ensuremath{\text{[\ion{O}{2}]}\lambda\lambda3726,3729}\xspace}
\newcommand{\oiione}{\ensuremath{\text{[\ion{O}{2}]}\lambda3726}\xspace}
\newcommand{\oiitwo}{\ensuremath{\text{[\ion{O}{2}]}\lambda3729}\xspace}
\newcommand{\oiinow}{\ensuremath{\text{[\ion{O}{2}]}\xspace}}

\newcommand{\neiiione}{\ensuremath{\text{[\ion{Ne}{3}]}\lambda3869}\xspace}

\newcommand{\niitot}{\ensuremath{\text{[\ion{N}{2}]}\lambda\lambda6548,6583}\xspace}
\newcommand{\niione}{\ensuremath{\text{[\ion{N}{2}]}\lambda6548}\xspace}
\newcommand{\niitwo}{\ensuremath{\text{[\ion{N}{2}]}\lambda6583}\xspace}

\newcommand{\siitot}{\ensuremath{\text{[\ion{S}{2}]}\lambda\lambda6717,6731}\xspace}
\newcommand{\siione}{\ensuremath{\text{[\ion{S}{2}]}\lambda6717}\xspace}

\newcommand{\hbeta}{\ensuremath{\text{H}\beta}\xspace}
\newcommand{\halpha}{\ensuremath{\text{H}\alpha}\xspace}
\newcommand{\lya}{\ensuremath{\text{Ly}\alpha}\xspace}

\newcommand{\ebv}{\ensuremath{E{(\mathit{B}-\mathit{V})}}\xspace}
\newcommand{\ebvstar}{\ensuremath{E{(\mathit{B}-\mathit{V})}_\text{star}}\xspace}
\newcommand{\ebvneb}{\ensuremath{E{(\mathit{B}-\mathit{V})}_\text{neb}}\xspace}

\newcommand{\uvbeta}{\ensuremath{\beta_\text{UV}}\xspace}

\newcommand{\ohmetal}{\ensuremath{12+\log(\text{O/H})}\xspace}

\newcommand{\autorefsec}[1]{\hyperref[#1]{Section~\ref*{#1}}}

\newcommand{\deltak}{\ensuremath{\Delta m_\mathit{Ks}}\xspace}

\newcommand{\fesc}{\ensuremath{f_\text{esc}}\xspace}
\newcommand{\xiion}{\ensuremath{\xi_\text{ion}}\xspace}
\newcommand{\xiionz}{\ensuremath{\xi_\text{ion,0}}\xspace}
\newcommand{\xiionunit}{\ensuremath{\text{erg}^{-1}\,\text{Hz}}\xspace}

\newcommand{\cigale}{\texttt{CIGALE}\xspace}

\definecolor{tomato}{rgb}{1.0, 0.39, 0.28}

\definecolor{orangered}{RGB}{255, 69, 0}

\graphicspath{{./}{figures/}}

\received{}
\revised{}
\accepted{}
\submitjournal{ApJ}

\shorttitle{Extreme emission line galaxies at $z \simeq 3.3$}
\shortauthors{Onodera et al.}

\begin{document}

\title{Broad-band selection, spectroscopic identification, and physical properties of a population of extreme emission line galaxies at $3<z<3.7${}\footnote{Based on data collected at Subaru Telescope, which is operated by the National Astronomical Observatory of Japan, with a proposal ID of S17A-049.}}

\correspondingauthor{Masato Onodera}
\email{monodera@naoj.org}

\author[0000-0003-3228-7264]{Masato Onodera}
\affiliation{Subaru Telescope, National Astronomical Observatory of Japan, National Institutes of Natural Sciences (NINS), 650 North A'ohoku Place, Hilo, HI 96720, USA}
\affiliation{Department of Astronomical Science, The Graduate University for Advanced Studies, SOKENDAI, 2-21-1 Osawa, Mitaka, Tokyo, 181-8588, Japan}

\author[0000-0003-4442-2750]{Rhythm Shimakawa}
\affiliation{National Astronomical Observatory of Japan, National Institutes of Natural Sciences (NINS), 2-21-1 Osawa, Mitaka, Tokyo, 181-8588, Japan}

\author[0000-0002-3560-1346]{Tomoko L. Suzuki}
\affiliation{Astronomical Institute, Tohoku University, 6-3, Aramaki, Aoba-ku, Sendai, Miyagi, 980-8578, Japan}
\affiliation{National Astronomical Observatory of Japan, National Institutes of Natural Sciences (NINS), 2-21-1 Osawa, Mitaka, Tokyo, 181-8588, Japan}
\affiliation{Kapteyn Astronomical Institute, University of Groningen, P.O. Box 800, 9700AV Groningen, The Netherlands}

\author[0000-0002-4937-4738]{Ichi Tanaka}
\affiliation{Subaru Telescope, National Astronomical Observatory of Japan, National Institutes of Natural Sciences (NINS), 650 North A'ohoku Place, Hilo, HI 96720, USA}

\author[0000-0002-6047-430X]{Yuichi Harikane}
\affiliation{Department of Physics and Astronomy, University College London, Gower Street, London WC1E 6BT, UK}
\affiliation{National Astronomical Observatory of Japan, National Institutes of Natural Sciences (NINS), 2-21-1 Osawa, Mitaka, Tokyo, 181-8588, Japan}

\author[0000-0002-9321-7406]{Masao Hayashi}
\affiliation{National Astronomical Observatory of Japan, National Institutes of Natural Sciences (NINS), 2-21-1 Osawa, Mitaka, Tokyo, 181-8588, Japan}

\author[0000-0002-2993-1576]{Tadayuki Kodama}
\affiliation{Astronomical Institute, Tohoku University, 6-3, Aramaki, Aoba-ku, Sendai, Miyagi, 980-8578, Japan}

\author[0000-0002-0479-3699]{Yusei Koyama}
\affiliation{Subaru Telescope, National Astronomical Observatory of Japan, National Institutes of Natural Sciences (NINS), 650 North A'ohoku Place, Hilo, HI 96720, USA}
\affiliation{Department of Astronomical Science, The Graduate University for Advanced Studies, SOKENDAI, 2-21-1 Osawa, Mitaka, Tokyo, 181-8588, Japan}

\author[0000-0003-2965-5070]{Kimihiko Nakajima}
\affiliation{National Astronomical Observatory of Japan, National Institutes of Natural Sciences (NINS), 2-21-1 Osawa, Mitaka, Tokyo, 181-8588, Japan}

\author{Takatoshi Shibuya}
\affiliation{Kitami Institute of Technology, 165 Koen-cho, Kitami, Hokkaido 090-8507, Japan}



\begin{abstract}%
    We present the selection, spectroscopic identification, and physical properties
    of extreme emission line galaxies (EELGs) at $3<z<3.7$
    aiming at studying physical properties of an analog population of
    star-forming galaxies (SFGs) at the epoch of reionization.
    The sample is selected based on the excess in the observed \textit{Ks} broad band flux
    relative to the best-fit stellar continuum model flux.
    By applying a 0.3 mag excess as a primary criterion,
    we select 240 EELG candidates with intense emission lines
    and estimated observed-frame equivalent width (EW) of $\gtrsim 1000$~\AA{}
    over the UltraVISTA-DR2 ultra-deep stripe in the COSMOS field.
    We then carried out a \textit{HK} band follow-up spectroscopy
    for 23 of the candidates with Subaru/MOIRCS,
    and find that 19 and two of them are at $z>3$ with intense \oiii emission,
    and \halpha{} emitters at $z\simeq 2$, respectively.
    These spectroscopically identified EELGs at $z\simeq 3.3$ show, on average,
    higher specific star formation rates (sSFR) than the star-forming main sequence,
    low dust attenuation of $\ebv \lesssim 0.1$~mag,
    and high $\oiii / \oiinow$ ratios of~$\gtrsim 3$.
    We also find that our EELGs at $z\simeq 3.3$ have
    higher hydrogen ionizing photon production efficiencies (\xiion)
    than the canonical value ($\simeq 10^{25.2}\,\text{\xiionunit}$),
    indicating that they are efficient in ionizing their surrounding interstellar medium.
    These physical properties suggest that they are low metallicity galaxies
    with higher ionization parameters and harder UV spectra than normal SFGs,
    which is similar to galaxies with Lyman continuum (LyC) leakage.
    Among our EELGs, those with the largest $\oiii/\oiinow$ and $\text{EW}(\oiii)$ values
    would be the most promising candidates to search for LyC leakage.
\end{abstract}



\section{Introduction} \label{sec:intro}

Cosmic reionization is one of the most dramatic events
in the course of the evolution of the universe.
Since star-forming galaxies (SFGs) at early cosmic epochs, $z \gtrsim 6$,
appear to dominate the reionization process
\citep[e.g.,][]{bouwens:2015,robertson:2015,matsuoka:2018},
understanding the nature, such as stellar masses ($M_\star$), star formation rates (SFRs),
metallicities, and ionization parameters 
of these objects provides important clues on
the ionizing sources responsible to reionize the universe
as well as the early phase of the galaxy formation and evolution.
These high-redshift SFGs have been selected via broad-band dropout
or narrow-band excess techniques
\citep[e.g.,][]{
    kashikawa:2006,
    bouwens:2008,
    ota:2010,
    ouchi:2010,
    ouchi:2018,
    ellis:2013,
    ono:2018},
many of which are inferred to be young ($\lesssim 100$~Myr),
low-mass ($M_\star \lesssim 10^{8\text{--}9}M_\odot$),
actively star-forming with a specific star formation rate
($\text{sSFR}\equiv\text{SFR}/M_\star$) of $\gtrsim 10$~Gyr$^{-1}$,
and metal-poor ($\ohmetal \lesssim 8$).
Due to such physical conditions,
their rest-frame optical emission lines such as \oiiitot and \halpha{}
are expected to be ubiquitously strong \citep[e.g.,][]{faisst:2016}
such that the inferred rest-frame equivalent width ($\text{EW}$) of $\hbeta+\oiiitot$
often exceeds $\gtrsim 1000$~\AA{}
\citep[e.g.,][]{smit:2014,smit:2015,robertsborsani:2016,stark:2017,debarros:2019,endsley:2020}.
However, currently available facilities both in space and on the ground
do not allow spectroscopic access to these emission lines at $z \gtrsim 4$,
preventing us from rest-frame optical emission line studies of
interstellar medium (ISM) properties as commonly done at $z \lesssim 4$.

Alternatively, lower-redshift analogs which are likely to have the similar characteristics
to high redshift galaxies have been sought.
In the local universe, these galaxies are called as green peas
\citep[GPs;][]{cardamone:2009, amorin:2010, izotov:2011}
named after their green colors due to extremely strong \oiii lines
(typically rest-frame $\text{EW} \gtrsim 100$~\AA{} and sometimes exceeds $\sim 1000$~\AA).
At intermediate redshifts of $1 \lesssim z \lesssim 3$,
the grism spectroscopic capability of \textit{HST}/WFC3
enables to find so-called ``Extreme Emission Line Galaxies'' (EELGs),
SFGs with similarly large emission line equivalent widths
\citep[e.g.,][]{straughn:2008, atek:2011, vanderwel:2011, maseda:2014}.
These studies indeed revealed the physical properties of EELGs
to be young, low-mass, low metallicity, and with high sSFR which is $\gtrsim 2\sigma$
above the average $M_\star$--SFR relation of SFGs at $z\simeq2$
\citep[e.g.,][]{whitaker:2014:ms}.

Although these previous works at $z<3$ have unveiled various properties of EELGs,
a detailed spectroscopic study of them as close redshifts
to the epoch of reionization (EoR) as possible is desirable
since the physical conditions of the universe can be significantly different at $z \gtrsim 3$.
In particular, the gas consumption timescale
would become similar or longer than the mass increase timescale at $z \gtrsim 3$
due to the increased matter accretion rate,
which likely causes a breakdown of a self-regulation of star formation in galaxies
\citep[e.g.,][]{lilly:2013}.
A steep decline ($\simeq 0.3$ dex) of the gas-phase metallicity of
normal SFGs from $z\simeq 2.3$ to $z \simeq 3.3$
presumably reflects the changes in the timescales
\citep[][but see \citealt{sanders:2020:mzr}]{maiolino:2008,onodera:2016,wuyts:2016}.
However, only a handful EELGs have spectroscopically
confirmed rest-frame emission line strengths at $z > 3$
and most of them are gravitationally lensed sources
\citep[e.g.,][]{fosbury:2003,amorin:2014,bayliss:2014,debarros:2016,cohn:2018}.
Therefore, a systematic study of EELGs at $z \gtrsim 3$ is still required to
obtain a comprehensive picture of the population itself
and to understand their higher redshift counterparts
(see also \citealt{tran:2020} for a recent observation of
rest-frame optical emission lines of EELGs at $3<z<3.8$).

In this study, we present a systematic search of EELGs at $3\lesssim z \lesssim 3.7$
with observed-frame $\text{EW}(\hbeta + \oiiitot)>1000\,\text{\AA}$
based on photometric data,
and their spectroscopic properties from our near-infrared (IR) follow-up observation.
The selection method and photometric properties of the sample are presented in
\autorefsec{sec:sample}.
Spectroscopic follow-up is presented in \autorefsec{sec:specprop},
and we derive physical parameters of the spectroscopically confirmed objects in \autorefsec{sec:prop}.
In \autorefsec{sec:results}, we show the results on the physical properties of
the identified EELGs at $z \simeq 3.3$ and discuss the implications for the
Lyman continuum (LyC) photon escape from these objects
and for properties of galaxies in the EoR.
We summarize our results in \autorefsec{sec:summary}.

Throughout the paper,
we adopt the WMAP 7 cosmology \citep{komatsu:2011:wmap7}
with $H_0=70.4\,\kms {\,}\mathrm{Mpc}^{-1}$, $\Omega_\text{m}=0.272$, and $\Omega_\Lambda=1 - \Omega_\text{m}$,
and AB magnitude system \citep{oke:1983}.

\section{Sample selection}\label{sec:sample}

\begin{figure}
    \centering
    \includegraphics[width=\linewidth]{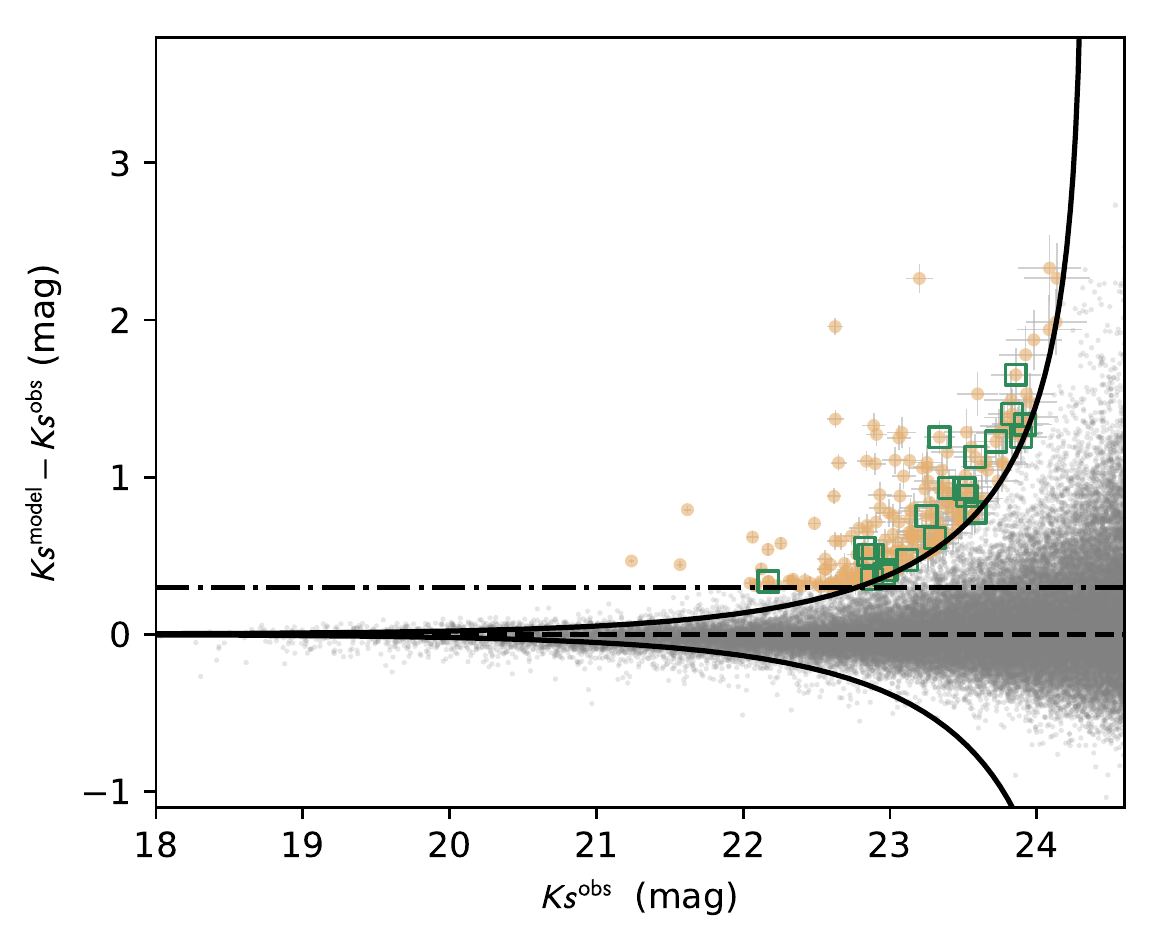}
    \caption{Color excess in \textit{Ks} band, $\deltak$,
        as a function of observed \textit{Ks}-band magnitude
        for all galaxies (dots) in the ultra-deep stripes of UltraVISTA-DR2.
        Beige circles with error bars highlight objects selected as EELG candidates
        satisfying the criteria $\deltak>0.3$ (dash-dotted line) and \Cref{eq:excess} (solid lines) on the positive side.
        Objects with spectroscopic identification by our Subaru/MOIRCS observation are shown with squares.
        Dashed line indicates the case of no excess.
        \label{fig:cmd}
    }
\end{figure}

We use the COSMOS2015 catalog \citep{laigle:2016} for the sample selection.
The COSMOS2015 catalog contains multiband photometric data
in the 2 deg$^{2}$ COSMOS field \citep{scoville:2007},
including \textit{Y}-band Hyper Suprime-Cam (HSC) image \citep{tanaka:2017:hsc_ssp_uh},
\textit{YJHKs} UltraVISTA-DR2\footnote{\url{www.eso.org/sci/observing/phase3/data_releases/uvista_dr2.pdf}} images \citep{mccracken:2012},
and $3.6\,\micron$ and $4.5\,\micron$ IR data
from the \textit{Spitzer} Large Area Survey with Hyper Suprime-Cam (SPLASH) project\footnote{\url{http://splash.caltech.edu/}}.
The goal of our photometric selection is to construct a parent sample of objects
at $3\lesssim z \lesssim 3.7$ with excess fluxes in the \textit{Ks}-band due to intense \hbeta and \oiiitot lines.

In order to estimate the contribution from the emission lines in the \textit{Ks}-band magnitude,
we first estimated stellar continuum by using the best-fit model spectral energy distribution (SED)
obtained by \texttt{EAZY}\footnote{\url{https://github.com/gbrammer/eazy-photoz}} \citep{brammer:2008}.
For the \texttt{EAZY} run, we fixed the redshifts to COSMOS2015 photometric redshifts
and used $3''$ aperture magnitudes in CFHT/MegaCam \textit{u}-band \citep{capak:2007},
Subaru/Suprime-Cam \textit{B}-, \textit{V}-, \textit{r}-, \textit{ip}-, and \textit{zpp}-band \citep{taniguchi:2007,taniguchi:2015},
VISTA/VIRCAM \textit{YJH}-band \citep{mccracken:2012},
and \textit{Spitzer}/IRAC $3.6\,\micron$ and $4.5\,\micron$ data (Moneti et al., in preparation).
We derived the stellar-only model \textit{Ks} magnitude, \textit{Ks}$^\mathrm{model}$,
by convolving the best-fit SED with the VISTA/VIRCAM \textit{Ks}-band transmission.
We then computed the difference between \textit{Ks}$^\mathrm{model}$ and the observed \textit{Ks} magnitudes, \textit{Ks}$^\mathrm{obs}$,
$\deltak \equiv \mathit{Ks}^\mathrm{model} - \mathit{Ks}^\mathrm{obs}$.
The \textit{Ks}-band excess magnitude is related to the observed-frame emission line equivalent widths as
\begin{equation}
    \deltak = -2.5 \log \frac{\mathrm{FWHM}_\mathit{Ks}}{\mathrm{EW}_\text{obs} (\oiii + \hbeta) +\mathrm{FWHM}_\mathit{Ks}},
    \label{eq:deltakew}
\end{equation}
where $\text{FWHM}_\mathit{Ks}$ is the band width of the VISTA \textit{Ks}-band filter \citep[3090~\AA;][]{laigle:2016}.
Our primary selection criterion is $\deltak>0.3$ corresponding to
the observed-frame $\mathrm{EW}>1000\,\mathrm{\AA}$ \citep{yamada:2005}.
At fainter magnitudes, one has to take the photometric errors into account.
Following \citet{bunker:1995}, we use the \textit{Ks}-band magnitude-dependent criteria,
\begin{equation}
    \deltak = -2.5 \log_{10} \left[ 1 - \frac{ \Sigma \sqrt{ f_{1\sigma_\mathrm{model}}^2 + f_{1\sigma_\mathrm{obs}}^2 }}{f_\mathrm{obs}} \right]
    \label{eq:excess}
\end{equation}
where
$f_\mathrm{obs}$,
$f_{1\sigma_\mathrm{obs}}$, and
$f_{1\sigma_\mathrm{model}}$
are an observed flux, $1\sigma$ error in the observed flux, and $1\sigma$ error in the model flux, respectively,
and $\Sigma$ sets the significance of the flux excess.
Here, we adopt $\Sigma=3$ and $f_{1\sigma_\mathrm{obs}}$ converted from
the $3\sigma$ depth of the ultra-deep stripes of UltraVISTA-DR2, 24.7~mag, as described in \citet{laigle:2016},
while we assume $f_{1\sigma_\mathrm{model}}=0$ for the model flux.

\Cref{fig:cmd} shows $\deltak$ as a function of the observed \textit{Ks}-band magnitude for
all galaxies in the ultra-deep stripes.
Objects satisfying $\deltak > 0.3$ and \Cref{eq:excess}
are shown with beige circles.
There are 240 such EELG candidates in the ultra-deep stripes of which 142 of them are $3 < z_\text{phot} < 3.7$.
Note that we only consider galaxies in a 0.46 deg$^2$ area of non-flagged regions both in UltraVISTA-DR2 and in the optical images \citep{capak:2007}.

\begin{figure*}
    \centering
    \includegraphics[width=0.95\linewidth]{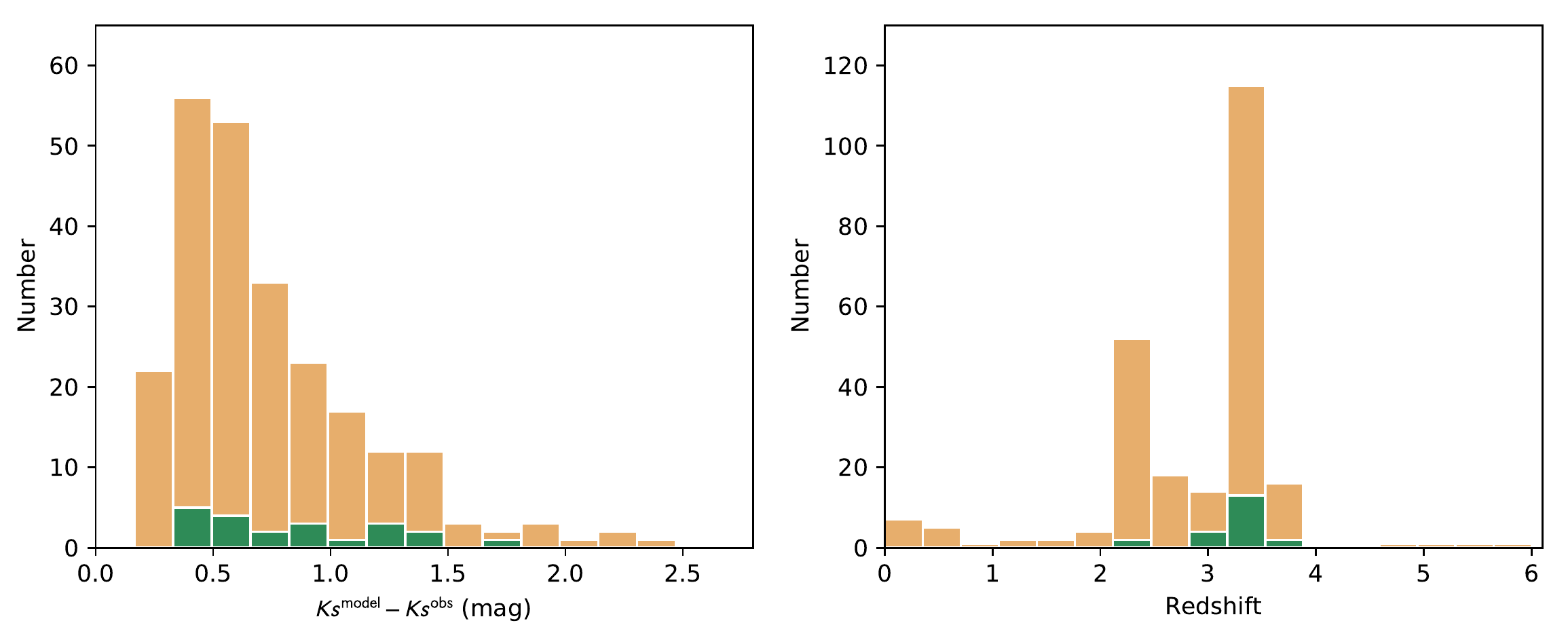}
    \caption{
        Histograms of $\deltak$ (\textit{Left}) and redshift (\textit{Right}) of
        240 EELG candidates selected in \autorefsec{sec:sample} (beige).
        The distributions of spectroscopically confirmed objects are
        overplotted with green histograms.
        \label{fig:phothist}
    }
\end{figure*}

The left panel of \Cref{fig:phothist} shows the distribution of \deltak of the EELG candidates as a beige histogram.
While the cut at 0.3 mag is our selection criteria,
the flux excess distribution shows a long tail reaching to $\sim 2.3$~mag,
corresponding to an observed EW of $\simeq 2.3 \times 10^4\,\mathrm{\AA}$
or a rest-frame $\mathrm{EW}(\hbeta + \oiiitot)\simeq 5300\,\mathrm{\AA}$ at $z \simeq 3.3$,
though part of such a large apparent excess can be due to
large photometric errors at fainter $\mathit{Ks}^\text{obs}$.
Most of the objects have $\deltak < 1.5$ corresponding to
the rest-frame $\mathrm{EW}(\hbeta + \oiiitot) \lesssim 3000\,\mathrm{\AA}$.
This can be translated into $\mathrm{EW}(\oiiitwo) \lesssim 2000\,\mathrm{\AA}$ \citep{reddy:2018:ew}.

The redshift distribution of the EELG candidates are shown in the right panel of \Cref{fig:phothist}.
Here photometric redshifts from the COSMOS2015 catalog are adopted.
EELG candidates with \textit{Ks}-band excess are mostly confined at $2 \lesssim z \lesssim 3.7$
with two notable peaks at $z \simeq 2.3$ and $z \simeq 3.4$,
corresponding to \halpha emitters and \oiii emitters, respectively.
A small number of objects (about 10~\%{}) is found outside this confined peaks.
Looking at their SEDs, the most likely explanation of these outliers is wrong photometric redshifts
and sometimes large photometric uncertainties in optical bands.

We have also carried out the same selection procedure for galaxies in the deep layer of UltraVISTA-DR2
by setting the $3\sigma$ \textit{Ks}-band magnitude limit as 24.0 and
found 318 objects satisfying the aforementioned criteria.
However, because of the shallower \textit{YJHKs} images of UltraVISTA-DR2 up to 0.7 mag,
photometric errors in these bands seem too large to robustly constrain
the continuum flux in SED fitting and the resulting selection appears to be significantly more uncertain than that in the ultra-deep stripes.
In order to keep the purity of the selection as much as possible,
we restrict our analysis and subsequent follow-up spectroscopy to the objects in the ultra-deep stripes.

\section{Spectroscopic follow-up observation}\label{sec:specprop}

\subsection{Follow-up near-IR spectroscopy}\label{sec:specobs}

We carried out a spectroscopic follow-up observation
for a subset of the \textit{Ks}-band excess EELG candidates.
Spectroscopic targets were selected primarily to maximize the number of
observed EELG candidates with $3 < z_\mathrm{phot} < 3.7$.
We also supplementarily considered criteria
$\mathit{rJL} \equiv (\mathit{J}-[3.6]) - 1.4(\mathit{r}-\mathit{J}) > 0$
which was proposed to select objects at $2.5 \lesssim z \lesssim 4 $
\citep{daddi:2004:bzk},
and \textit{iJHK} colors to remove low redshift ($z \lesssim 0.2$) emission line galaxies \citep{tadaki:2013}.

We used Multi-Object InfraRed Camera and Spectrograph \citep[MOIRCS;][]{ichikawa:2006:moircs,suzuki:2008:moircs}
at the Subaru Telescope.
The two detectors of MOIRCS have been replaced from Hawaii-2 to Hawaii-2RG in 2015 \citep{fabricius:2016:moircs,walawender:2016:moircs}.
We used the HK500 grism covering $1.3\,\micron$ to $2.3\,\micron$ with a $0.8$~arcsec slit which provides the spectral resolution of $R\simeq 350$.
After the detector upgrade, the lower readout noise allows us to shorten the exposure time to reach the background limited performance,
enabling us to better capture the temporal variation of the sky background.
Although the throughput in the \textit{HK}-band improved only by $\lesssim 5$~\%{},
the lower readout noise results in a deeper limiting flux than the previous detectors
by cleaner sky subtraction, which is crucial for the relatively low resolution of the HK500 grism used in this study.

The observation was carried out during the first half nights on April 9 and 17, 2017
under clear sky condition with 0.5--0.8~arcsec seeing.
Three masks for three pointings (each has a diameter of $\simeq 6$~arcmin) were used to observe in total 23 EELG candidates.
We used a modified AB dithering pattern with three dithering widths of 2.8, 3.0, and 3.2~arcsec
to increase the signal-to-noise ratio (S/N) and to avoid bad pixels \citep{kriek:2015:mosdef},
and 180~s on-source exposure per dithering position.
Each mask has been integrated $\simeq 40$--$110$~minutes.
The detailed observing log is shown in \Cref{tab:obslog}.
We observed A0V-type stars at the beginning of the nights
for flux calibration as well as telluric correction
using the identical instrument setup to the science exposures.

Data reduction was carried out using the MCSMDP\footnote{\url{http://www.edechs.com/MCSMDP/}} pipeline \citep{yoshikawa:2010} and custom scripts.
Since the original MCSMDP was developed for the previous detectors,
we made modifications to properly handle the new Hawaii-2RG detectors.
The data were flat-fielded by using dome-flat frames which were taken under the identical
instrument setup to the science frames.
Bad pixels and cosmic rays were identified by using bad-pixel maps
created for the new detectors and by using the pair of images in the dithering, respectively,
and linearly interpolated using adjacent pixels along the spatial direction.
Background sky was subtracted by taking a difference between A and B positions.
The optical distortion was then corrected by applying a polynomial function
determined by the MOIRCS imaging reduction pipeline
MCSRED\footnote{\url{https://www.naoj.org/staff/ichi/MCSRED/mcsred_e.html}}.

Individual 2-dimensional (2D) spectra were extracted from each frame
and the wavelength calibration was carried out
by using OH-airglow lines \citep{rousselot:2000}.
The typical uncertainty of the wavelength calibration is $\simeq 3$~\AA{} which is about
40~\%{} of the pixel scale.
Based on the wavelength solution, the extracted 2D spectra were rectified
so that sky lines are aligned along the spatial direction
and the secondary sky subtraction procedure was made in order to
remove residual background by fitting a linear function at each wavelength pixel.

Standard star frames of A0V-type stars were processed in the same manner as the science object frames.
The 1D spectra of the standard stars were extracted using the \texttt{apall} task in
Image Reduction and Analysis Facility \citep[IRAF;][]{tody:1986:iraf,tody:1993:iraf}
and system total response curve was derived by comparing the observed spectra
with a theoretical template \citep{kurucz:1979}.
Along with the telluric correction,
we have carried out absolute flux calibration for the standard spectra
by scaling them to 2MASS magnitudes correcting for the slit loss.
Flux calibrated 2D frames of each object are then median stacked after $3\sigma$ clipping
with offsets corresponding to the dithering widths.
One dimensional noise spectra were calculated using the $1\sigma$ dispersion
of the counts along the spatial direction at each wavelength pixel
in the non-detected slits for each mask after the stacking procedure.

One dimensional science and corresponding noise spectra were extracted
by adopting the optimal extraction algorithm \citep{horne:1986}.
Here, we assume a Gaussian spatial profile at the strongest detected emission line for weighting
and a straight trace for extraction.
Before the extraction, we confirmed that the trace is straight along the dispersion direction
by using bright stars and galaxies put in the science masks as filler targets.

\begin{figure*}
    \centering
    \includegraphics[width=\linewidth]{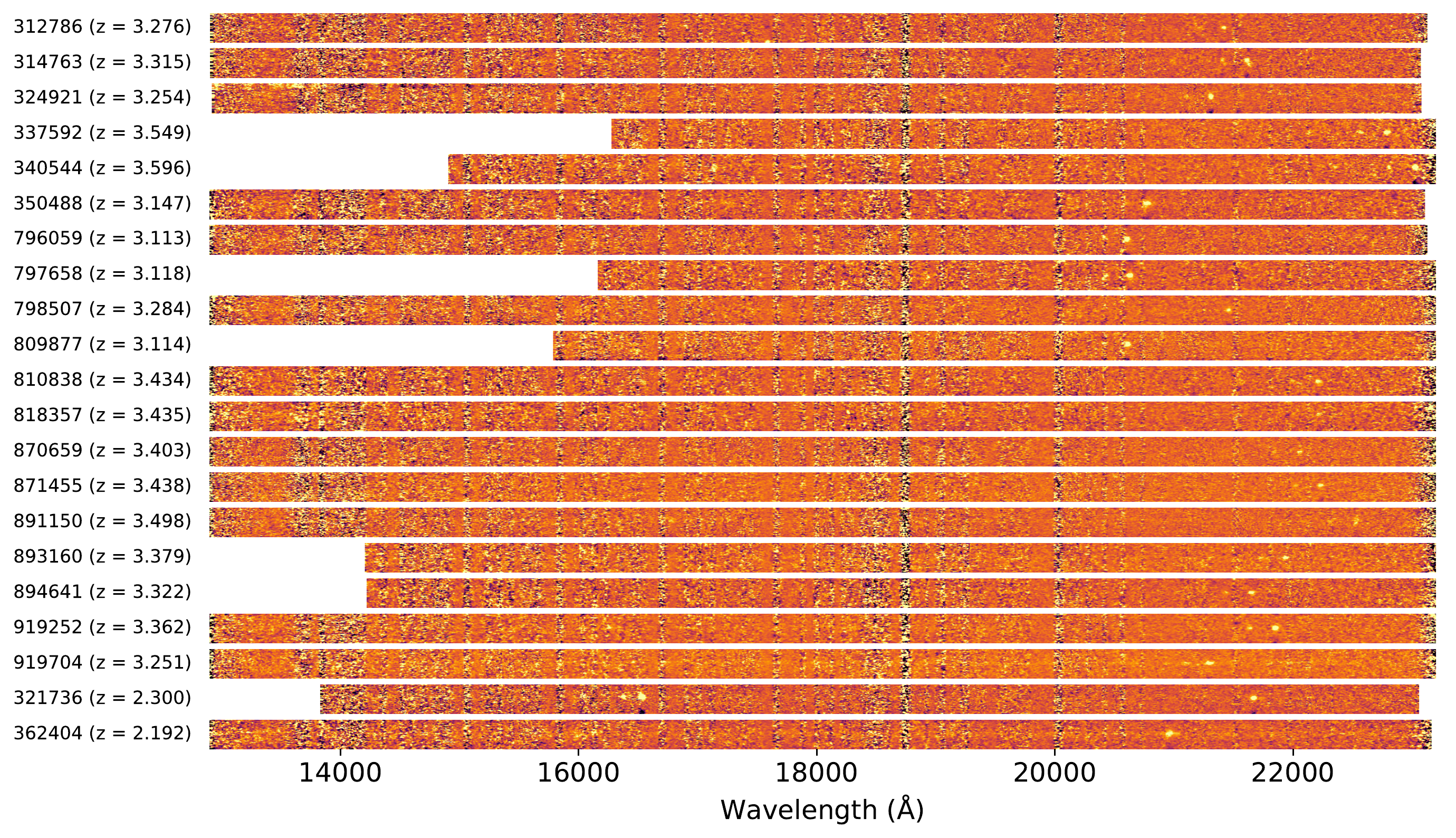}
    \caption{
        Two-dimensional Subaru/MOIRCS spectra of 21 EELG candidates with detected emission lines.
        Object IDs and spectroscopic redshifts are indicated on the left side of the spectra.
        \label{fig:spec2d}
    }
\end{figure*}

\begin{figure*}
    \centering
    \includegraphics[width=\linewidth]{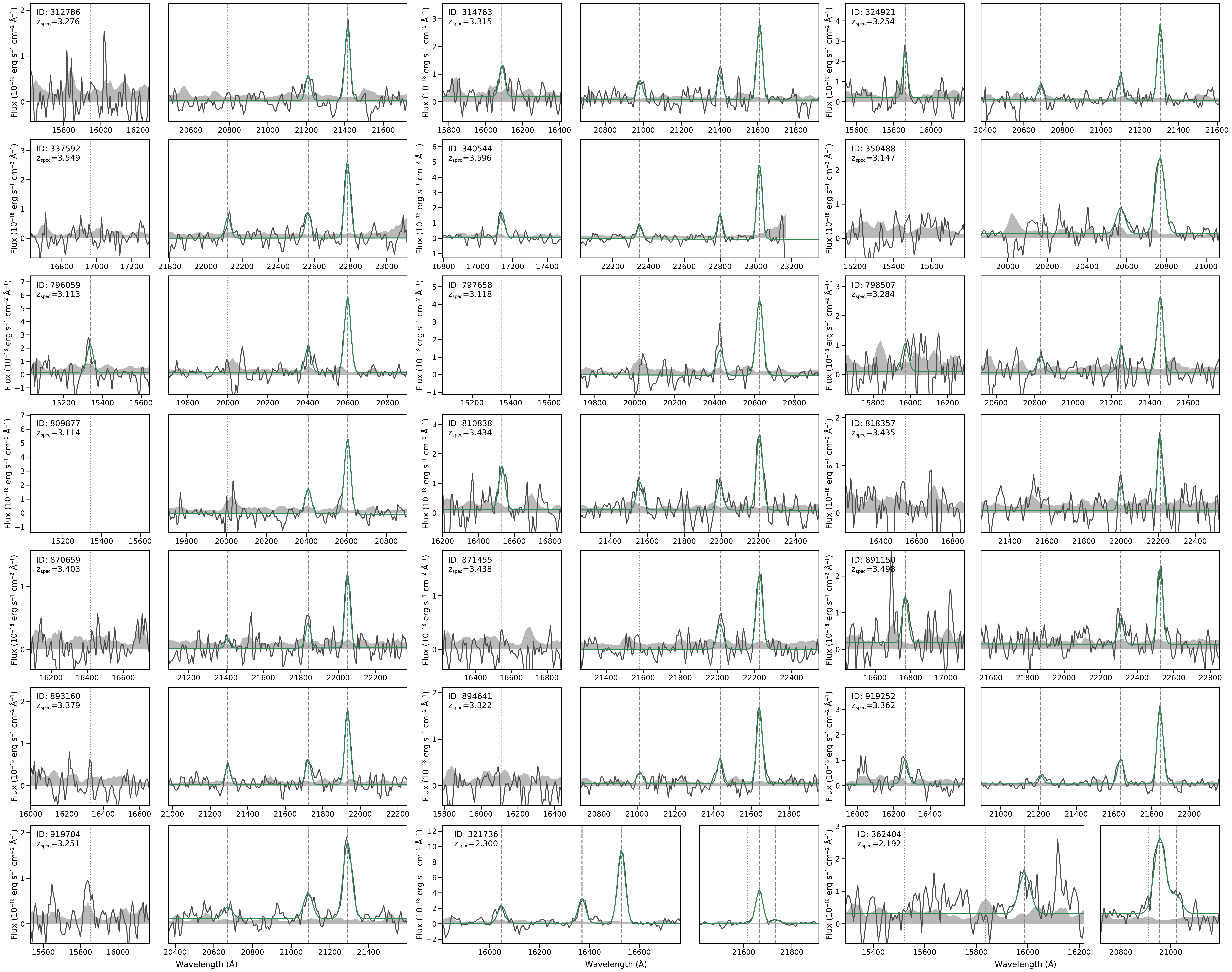}
    \caption{
        One-dimensional Subaru/MOIRCS spectra of 21 EELG candidates with detected emission lines.
        Object IDs and spectroscopic redshifts are indicated on the left side of the spectra.
        Spectra of each object are shown in a pair of panels
        corresponding to \textit{H} and \textit{K} bands.
        In each panel, object and noise spectra are shown with
        solid black lines and gray shaded areas, respectively,
        together with the best-fit Gaussian functions and continuum in green solid lines.
        Vertical lines show the locations of detected (dashed) and undetected (dotted) emission lines.
        These emission lines corresponds to \oii, \hbeta, and \oiiitot for the first 19 objects,
        and \hbeta, \oiiitot, \niione, \halpha, and \niitwo for the last two objects, from left to right.
        \label{fig:spec1d}
    }
\end{figure*}

\Cref{fig:spec2d} and \Cref{fig:spec1d}, respectively,
shows 2D and 1D spectra of emission line detected EELG candidates.
Twenty-one out of 23 observed EELG candidates show detected emission lines
and the origin of \textit{K}-band excesses are confirmed
as \oiii and \halpha at $z\simeq 3.3$ and $z\simeq 2.2$ for 19 and 2 objects, respectively.

\begin{deluxetable}{lcccccc}
  \tablecaption{
    EELG candidates for the near-IR spectroscopic follow-up with Subaru/MOIRCS
    \label{tab:obslog}
  }
  \tablehead{
    \colhead{ID} &
    \colhead{R.A.} &
    \colhead{Decl.} &
    \colhead{$z_\mathrm{phot}$} &
    \colhead{$\mathit{Ks}_\mathrm{tot}$} &
    \colhead{$\Delta m_\mathit{Ks}$} &
    \colhead{$T_\mathrm{exp}$} \\
    \colhead{ } &
    \colhead{(deg)} &
    \colhead{(deg)} &
    \colhead{ } &
    \colhead{(mag)} &
    \colhead{(mag)} &
    \colhead{(minutes)} \\
    \colhead{(1)} &
    \colhead{(2)} &
    \colhead{(3)} &
    \colhead{(4)} &
    \colhead{(5)} &
    \colhead{(6)} &
    \colhead{(7)}}
  \startdata{}
  312786 & 149.77668 & 1.763990 & 3.326 & 23.50 & 0.93 & 108 \\
  314763 & 149.74508 & 1.766926 & 3.520 & 22.85 & 0.55 & 108 \\
  324921 & 149.73393 & 1.782772 & 3.464 & 22.91 & 0.38 & 108 \\
  337592 & 149.78500 & 1.802777 & 3.555 & 23.87 & 1.65 & 108 \\
  340544 & 149.79346 & 1.807244 & 3.524 & 23.61 & 0.78 & 108 \\
  350488 & 149.77386 & 1.822980 & 3.422 & 23.04 & 0.41 & 108 \\
  796059 & 150.42306 & 2.509079 & 3.316 & 22.86 & 0.51 & \phn{}42 \\
  797658 & 150.47565 & 2.511852 & 3.320 & 23.59 & 1.13 & \phn{}42 \\
  798507 & 150.44682 & 2.512701 & 3.374 & 23.31 & 0.75 & \phn{}42 \\
  809877 & 150.46558 & 2.529429 & 3.166 & 23.36 & 1.25 & \phn{}42 \\
  810838 & 150.50135 & 2.531063 & 3.500 & 23.34 & 0.61 & \phn{}42 \\
  818357 & 150.50653 & 2.542366 & 3.402 & 23.99 & 1.26 & \phn{}42 \\
  870659 & 149.80901 & 2.622814 & 3.416 & 23.77 & 1.40 & \phn{}84 \\
  871455 & 149.79484 & 2.623711 & 3.407 & 23.56 & 0.88 & \phn{}84 \\
  891150 & 149.77483 & 2.653322 & 3.541 & 22.88 & 0.35 & \phn{}84 \\
  893160 & 149.84076 & 2.656873 & 3.316 & 24.00 & 1.34 & \phn{}84 \\
  894641 & 149.83600 & 2.659291 & 3.354 & 23.71 & 1.23 & \phn{}84 \\
  919252 & 149.82153 & 2.696576 & 3.292 & 23.36 & 0.93 & \phn{}84 \\
  919704 & 149.79926 & 2.697392 & 3.249 & 22.88 & 0.50 & \phn{}84  \\
  321736 & 149.79515 & 1.777986 & 2.300 & 23.13 & 0.47 & 108 \\
  362404 & 149.74646 & 1.841587 & 2.199 & 22.13 & 0.34 & 108 \\
  348570\tablenotemark{a} & 149.76113 & 1.819842 & 3.774 & 23.42 & 0.73 & 108 \\
  815834\tablenotemark{a} & 150.44337 & 2.538413 & 2.476 & 23.00 & 0.52 & \phn{}42
  \enddata{}
  \tablecomments{
    (1) Object ID \citep{laigle:2016};
    (2) R.A.;
    (3) Decl;
    (4) photometric redshift from \citet{laigle:2016};
    (5) total \textit{Ks} magnitude;
    (6) $\Delta m_\mathit{Ks}$;
    (7) exposure time in Subaru/MOIRCS spectroscopic observation.
  }
  \tablenotetext{a}{Objects without any emission line detections.}
\end{deluxetable}

\subsection{Emission line measurement}\label{sec:specmeas}

As seen in \Cref{fig:spec2d} and \Cref{fig:spec1d},
21 objects show significant detection of multiple emission lines,
which makes the emission line identification and rough redshift estimate straightforward.
Then we measured the emission line properties in two fitting processes as explained below.

First, we only fit the primary emission line.
The primary emission line here is the strongest emission line,
i.e., \oiiitwo and \halpha for $z\simeq3.3$ and $z\simeq 2.2$, respectively.
We assume a Gaussian profile with a flat continuum.
Therefore, there are four free parameters,
namely, the redshift, line width $\sigma$, total line flux, and constant continuum flux.
We used $\pm 500$~\AA{} from the line center for the fitting procedure.
When the primary line is \halpha,
adjacent \niitot lines are fit simultaneously as they are not fully resolved.
In this case, all 3 lines are assumed to have the same redshift, $\sigma$, and continuum flux.
Also, \niione flux is assumed to be one third of \niitwo flux.

For the fitting, we used \texttt{emcee}, a Python implementation of an affine invariant Markov chain Monte Carlo (MCMC) algorithm
\citep{emcee,goodman:2010} through \texttt{lmfit} \citep{lmfit}.
Priors are assumed to be top-hat in all free parameters with only minimal boundaries, namely positivity of the $\sigma$ and line flux.
We used 100 walkers and 1000 steps per walker of which 200 steps were used as a burn-in process
and discarded from the posterior sampling.
Since the S/N ratios are typically high $\gg 10$ for the primary emission lines,
the convergence is quickly achieved and all parameters for line profiles are well constrained.

In the second fitting process, we fit the rest of strong emission lines,
namely, \oiitot, \neiiione, \hbeta, and \siitot
within the observed wavelength range.
For $z\simeq2.2$ objects, \oiiitot were also fit in this process.
Here, the redshift and line width $\sigma$ was fixed to
those determined in the first fitting process,
but allowing to be different within $\pm 0.5\sigma$ and $\pm 0.1\sigma$
from those of the primary line, respectively.
We also used $\pm 500$~\AA{} from the line centers for the fitting,
but continuum was assumed to be a linear function across the spectrum.
Because non-primary lines are fainter than the primary and include non-detections,
we used 100 walkers with 10000 steps per walker and 2000 burn-in steps
to sample the posterior distributions.
These numbers are large enough to distinguish between converged and non-converged parameters.

For each parameter, the best fit parameter and the corresponding $1\sigma$ error are
defined as the median and the half the interval between 16 and 84 percentiles, respectively,
of the posterior distribution.
To judge whether the line is detected or not,
we compared the best-fit line flux with the $3\sigma$ noise
computed from 1D noise spectra integrated with a Gaussian weight corresponding to the primary's line profile.
While continuum fluxes are not well constrained in general for our sample as
they are barely detected as seen in \Cref{fig:spec2d} and \Cref{fig:spec1d},
this does not affect the emission line flux measurements.
\Cref{tab:emfit} shows the measured emission line properties.
Here we only list emission line fluxes used in the following discussion,
though we also fit \neiiione and \siitot together in the second fitting process.
As a reference, there are 5 out of 19 and 1 of 2 objects at $z\simeq 3.3$ and $z=2.2$
with detected \neiiione and \siione at $>3\sigma$ significance, respectively.

In the following sections, we focus on 19 spectroscopically confirmed EELGs at $3.1 < z < 3.6$
unless explicitly mentioned.
Their positions are highlighted in \Cref{fig:cmd} and
the distributions of \deltak and spectroscopic redshifts are
shown in \Cref{fig:phothist}.
The median spectroscopic redshift of them is 3.3.

\begin{deluxetable*}{ccccccc}
  \tablecaption{
    Emission line measurements%
    \label{tab:emfit}
  }
  \tablehead{
    \colhead{ID} &
    \colhead{$z_\mathrm{moircs}$} &
    \colhead{$F(\oii)$} &
    \colhead{$F(\hbeta)$} &
    \colhead{$F(\oiiitwo)$} &
    \colhead{$F(\halpha)$} &
    \colhead{$F(\niitwo)$} \\
    \colhead{(1)} &
    \colhead{(2)} &
    \colhead{(3)} &
    \colhead{(4)} &
    \colhead{(5)} &
    \colhead{(6)} &
    \colhead{(7)}
  }
  \startdata
  \sidehead{\oiii{} emitters\tablenotemark{a}}
  312786 & $3.2758$ & $<  1.27$ & $<  0.69$ & $ \phn{}5.59 \pm  0.27$ & \nodata{} & \nodata{} \\
  314763 & $3.3148$ & $ 4.15 \pm  0.80$ & $ 2.66 \pm  0.27$ & $ \phn{}9.83 \pm  0.42$ & \nodata{} & \nodata{} \\
  324921 & $3.2539$ & $ 7.52 \pm  1.19$ & $ 2.52 \pm  0.35$ & $12.81 \pm  0.46$ & \nodata{} & \nodata{} \\
  337592 & $3.5492$ & $<  1.65$ & $ 2.84 \pm  0.48$ & $10.41 \pm  0.42$ & \nodata{} & \nodata{} \\
  340544 & $3.5965$ & $ 6.23 \pm  0.52$ & $ 3.29 \pm  0.35$ & $17.95 \pm  0.74$ & \nodata{} & \nodata{} \\
  350488 & $3.1468$ & $<  2.36$ & $<  1.89$ & $12.76 \pm  0.57$ & \nodata{} & \nodata{} \\
  796059 & $3.1132$ & $ 8.57 \pm  1.79$ & $<  6.23$ & $22.75 \pm  0.89$ & \nodata{} & \nodata{} \\
  797658 & $3.1181$ & \nodata{} & $<  5.76$ & $18.49 \pm  0.55$ & \nodata{} & \nodata{} \\
  798507 & $3.2837$ & $ 4.10 \pm  0.97$ & $ 2.35 \pm  0.51$ & $10.89 \pm  0.59$ & \nodata{} & \nodata{} \\
  809877 & $3.1144$ & \nodata{} & $<  6.45$ & $21.12 \pm  0.73$ & \nodata{} & \nodata{} \\
  810838 & $3.4338$ & $ 6.57 \pm  0.67$ & $ 4.24 \pm  0.64$ & $11.06 \pm  0.59$ & \nodata{} & \nodata{} \\
  818357 & $3.4348$ & $<  1.61$ & $<  1.52$ & $ \phn{}5.53 \pm  0.49$ & \nodata{} & \nodata{} \\
  870659 & $3.4028$ & $<  1.07$ & $ 0.59 \pm  0.18$ & $ \phn{}4.17 \pm  0.32$ & \nodata{} & \nodata{} \\
  871455 & $3.4381$ & $<  1.15$ & $<  0.89$ & $ \phn{}6.09 \pm  0.37$ & \nodata{} & \nodata{} \\
  891150 & $3.4976$ & $ 5.26 \pm  0.44$ & $<  1.62$ & $ \phn{}8.01 \pm  0.66$ & \nodata{} & \nodata{} \\
  893160 & $3.3792$ & $<  1.36$ & $ 1.67 \pm  0.18$ & $ \phn{}6.38 \pm  0.31$ & \nodata{} & \nodata{} \\
  894641 & $3.3216$ & $<  1.99$ & $ 0.94 \pm  0.15$ & $ \phn{}6.20 \pm  0.24$ & \nodata{} & \nodata{} \\
  919252 & $3.3617$ & $ 4.14 \pm  0.54$ & $ 1.33 \pm  0.25$ & $11.93 \pm  0.38$ & \nodata{} & \nodata{} \\
  919704 & $3.2512$ & $<  2.87$ & $ 1.46 \pm  0.25$ & $ \phn{}9.41 \pm  0.32$ & \nodata{} & \nodata{} \\
  \sidehead{\halpha{} emitters\tablenotemark{b}}
  321736 & $2.3002$ & \nodata{} & $ 8.60 \pm  1.16$ & $36.35 \pm  0.70$ & $15.72 \pm  0.39$ & $ 1.69 \pm  0.32$ \\
  362404 & $2.1923$ & \nodata{} & $<  2.96$ & $ \phn{}7.05 \pm  1.29$ & $13.19 \pm  0.47$ & $ 3.07 \pm  0.47$ \\
  \sidehead{Composite of \oiii{} emitters}
  Low-mass composite & \nodata{} & $1.70 \pm 0.13$ & $1.22 \pm 0.06$ & $\phn{}9.01 \pm 0.08$ & \nodata{} & \nodata{} \\
  High-mass composite & \nodata{} & $3.16 \pm 0.17$ & $1.46 \pm	0.09$ & $\phn{}9.79 \pm 0.11$ & \nodata{} & \nodata{}
  \enddata{}
  \tablecomments{
    (1) Object ID{};
    (2) spectroscopic redshift measured from MOIRCS spectra;
    (3) \oii{} ($\oiione + \oiitwo$) flux;
    (4) \hbeta{} flux;
    (5) \oiiitwo{} flux;
    (6) \halpha{} flux;
    and
    (7) \niitwo{} flux.
    All fluxes are in units of $10^{-17}\,\text{erg}\,\text{s}^{-1}\,\text{cm}^{-2}$,
    not corrected for dust extinction and stellar absorption.
    Quoted upper limits are the $3\sigma$ upper limit.
  }
  \tablenotetext{a}{$\oiiitwo$ is the primary emission line.}
  \tablenotetext{b}{$\halpha$ is the primary emission line.}
\end{deluxetable*}

\section{Measurement of physical properties}\label{sec:prop}

\subsection{Broad-band SED fitting} \label{sec:mass}

We carried out a SED fitting to broad-band photometry
to obtain primarily stellar masses.
Subaru/HSC \textit{Y} band data are also used
in addition to the the photometric bands used to derive \deltak.
Photometry is converted to the total magnitudes from 3 arcsec aperture magnitudes
following Appendix A.2 in \citet{laigle:2016}.
Upper limits are assigned when the fluxes are below $3\sigma$ significance.

SED fitting was performed by using the 2018.0 version of
Code for Investigating GALaxy Emission \citep[\cigale\footnote{\url{https://cigale.lam.fr/}}; ][]{burgarella:2005, noll:2009, boquien:2019}.
As SED templates, we used composite stellar population models generated
from the simple stellar population models of \citet[hereafter BC03]{bruzual:2003}
with a Chabrier initial mass function \citep[IMF;][]{chabrier:2003}
with lower and upper mass cutoffs of $0.1\,M_\odot$ and $100\,M_\odot$, respectively.
Stellar population age ranges in $\log\text{age}/\text{yr}=7$--$9.6$ with steps of 0.1 dex.
The upper limit of the age is assumed not to exceed the age of the universe at $z=3$.
Metallicities are allowed to have $Z = 0.004$, $0.008$, and $0.02$.

We consider delayed-$\tau$ models, $\text{SFR} \propto t \exp(-t/\tau)$,
for the star formation history (SFH)
with $\log\tau/\text{yr}=8$--$10$ with steps of 0.1 dex.
Recent studies of SFH of EELGs at low and high redshifts
suggest that they are in a starburst phase.
\citet{telles:2018} studied SFH of local \ion{H}{2} galaxies with a three-burst SFH
and found that the bulk of stellar mass of them has been formed by the past star formation
episode, while they are currently in a maximum starburst phase.
At higher redshifts similar to our sample,
\citet{cohn:2018} found that EELGs at $2.5 < z < 4$ show evidence of a starburst
in the most recent 50~Myr with rising SFH in the last 1~Gyr.
These results suggests that delayed-$\tau$ models are
more representative SFH for EELGs than a constant and exponentially declining SFHs.
Note that in the later analysis, we will only use the stellar mass from the SED fitting and
the stellar mass is a robust parameter against the assumption of SFH{}.

For intense emission line galaxies like the EELGs studied here,
inclusion of emission lines has critical importance
\citep[e.g.,][]{schaerer:2013,stark:2013,onodera:2016}
to extract physical parameters from SED fitting.
For example, changes in stellar masses can be $\gtrsim 0.5$ dex
when the contribution of $\hbeta + \oiiitot$ fluxes is $\gtrsim 50$~\%{}
in the \textit{Ks} band \citep{onodera:2016}.
Instead of subtracting emission line contributions from broad-band photometry \citep{onodera:2016},
we include nebular emissions as supported in \cigale.
We assumed an ionization parameter of $\log U=-2.5$
which is a typical value for star-forming galaxies on the star-forming main sequence (MS)
at $z\sim3.3$ \citep{onodera:2016}.
LyC photons are assumed to be entirely absorbed by neutral hydrogen,
i.e., zero LyC escape fraction ($\fesc$) and no LyC absorption by dust.
As we will discuss later, some of our EELGs at $z\simeq3.3$ are likely to have non-zero \fesc.
However, the photometric data from the COSMOS2015 catalog
do not allow us to put meaningful constraints on the \fesc with \cigale.
Indeed, the output parameters used in the following discussion
are consistent well within the corresponding errors even if we set \fesc as a free parameter.
Therefore, we simply adopt the fitting results with $\fesc=0$.
Nebular contributions (i.e., emission lines and continuum) are
calculated following \citet{inoue:2011} in \cigale.
Although the line width is not important for broad-band data,
we set FWHM of emission line as 300 \kms.

In \cigale dust attenuation is implemented as a modified starburst law, $k_\lambda$,
which is based on the Calzetti curve \citep{calzetti:2000}, $k_{\lambda}^{\text{starburst}}$,
with flexibilities to add an UV bump and alter the overall slope as follows.
\begin{equation}
    k_\lambda = D_\lambda + k^{\text{starburst}}_{\lambda} \left(\frac{\lambda}{550\,\text{nm}}\right)^{\delta}
    \frac{E (\mathit{B}-\mathit{V})_{\delta=0}}{E (\mathit{B}-\mathit{V})_\delta},
\end{equation}
where $D_\lambda$ is the Drude profile to express an UV bump
and $\delta$ modifies the slope \citep{noll:2009}.
We assumed the average SMC Bar extinction curve from \citet{gordon:2003}
which can be approximated by setting $D_\lambda = 0$ and $\delta = -0.62$.
The nebular extinction \ebvneb is allowed to vary between 0 to 0.8 with steps of 0.01 in the fitting.
Because nebular emission lines are explicitly taken into account in the fit,
one needs to make further assumptions on the attenuation curve and
the relation between attenuation affecting stellar continuum and nebular emission.
Here we also used the same SMC Bar curve \citep{gordon:2003} for nebular emission lines
and assumed $\ebvneb = 3.06\,\ebvstar$.
The latter relationship has been recently obtained by \citet{theios:2019}
for the case of the SMC attenuation curve
using a sample of star-forming galaxies at $2<z<2.7$ \citep{steidel:2014,strom:2017}.

At $z \gtrsim 3$, absorption by the intergalactic medium (IGM) in the observed optical bands
cannot be ignored \citep[e.g.][]{madau:1995,inoue:2008}.
When shifting the template SEDs to the observed redshift, \cigale applies
the IGM absorption using the prescription of \citet{meiksin:2006}.

\begin{figure*}
    \centering
    \includegraphics[width=\linewidth]{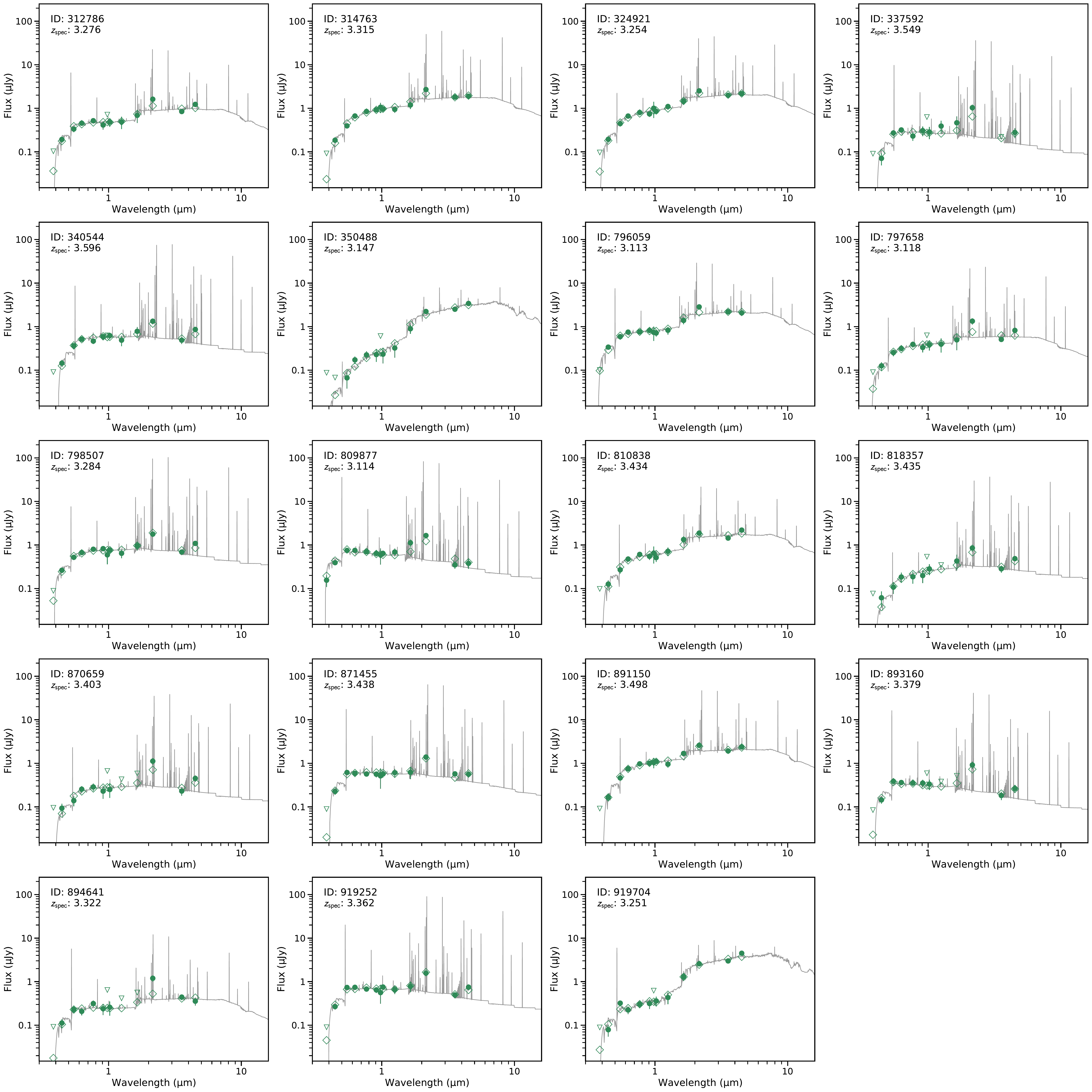}
    \caption{
        Broad-band SEDs of 19 spectroscopically identified EELG candidates at $z\simeq3.3$.
        Filled circles and open triangles show observed fluxes and $3\sigma$ upper limits, respectively.
        Best-fit SED templates derived by \cigale{} are shown as solid lines.
        Open diamonds are the best-fit SEDs convolved by the filter transmission curves used in the fit.
        Objects IDs and spectroscopic redshifts are shown at the upper left corner of each panel.
        \label{fig:sedfitting}
    }
\end{figure*}

The best-fit values and their corresponding standard deviations
are computed based on probability distribution functions (PDFs) implemented
as the \texttt{pdf\_analysis} module (see \citealt{noll:2009} and \citealt{boquien:2019} for the details).
The observed and best-fit SEDs are shown in \Cref{fig:sedfitting} and
the resulting stellar masses, UV spectral slopes (see \autorefsec{sec:uvprop}),
attenuation parameters, SFRs, and stellar ages for EELGs at $z>3$ are listed in \Cref{tab:sedprop}.

As seen in \Cref{fig:sedfitting}, most of EELGs show a flat or power-law continuum
in the optical to near-IR bands corresponding to young stellar populations
with an excess flux in the \textit{K} band due to the intense \oiiitwo line.
There are a couple objects, namely 350488 and 919704,
showing redder SEDs than the others.
These two objects indeed have the best-fit age of $\gtrsim 1$~Gyr,
indicating a significant contribution of old stellar populations.

It turned out that \cigale underestimates \oiiitwo fluxes and equivalent widths
of the EELG sample on average by a factor of two in our fitting run.
To check this effect on the estimate of the SED properties,
especially stellar masses,
we ran \cigale without \textit{Ks}-band data while keeping other inputs identical,
and found that stellar masses from the no-\textit{Ks} run are
larger than those from our fiducial run up to $0.2$~dex
with a median of $0.04$~dex.
The difference tends to be larger at lower stellar masses ($\log M_\star / M_\odot \lesssim 9$)
with a median of $0.08$~dex.
These differences are, however, smaller than the uncertainties
of stellar masses and the results and discussion presented later
will remain unchanged.

The underestimated best-fit predicions of \oiii fluxes and equivalent widths
could be an indication of higher ionization parameters in EELGs at $z\simeq 3.3$
than $\log U=-2.5$ assumed in the \cigale{} run.
Although $\log U > -2.0$ seems unrealistically high for SFGs
\citep[e.g.,][]{yeh:2012, shirazi:2014, strom:2018},
we carried out a \cigale{} run with a fixed $\log U=-1$ which
is a theoretical maximum \citep{yeh:2012}.
It is found, however, that the predicted \oiiitwo properties
are still underestimated, though the discrepancies become smaller.
Therefore, the issue cannot be solved solely by the higher ionization parameter.

The lower limit of the stellar population age in the SED fitting
could also limit the \oiiitwo properties.
To check this possibility, we run \cigale by adding $1$--$10$~Myr stellar populations.
We found that the best-fit $\text{EW} (\oiiitwo)$ does not always become larger by including
younger stellar populations, but a clear improvement is seen
for those with the highest $\text{EW} (\oiiitwo)$
and the best-fit ages of $<10$~Myr are often derived for such cases.
Similar to other \cigale runs discussed above,
there is only little change in the best-fit stellar mass ($0.01$~dex in median).
Trying to find the exact match of the equivalent widths and fluxes of the \oiiitwo emission line
from our EELGs sample is not a focus of this study,
but it appears that higher ionization parameters and younger stellar population ages
are likely to play key roles to intensify the emission line strength
(see also \autorefsec{sec:eor}; \citealt{debarros:2019}).

\begin{deluxetable*}{ccccccccc}
  \tablecaption{
    Stellar mass, attenuation, star formation rate, and stellar age
    \label{tab:sedprop}
  }
  \tablehead{
    \colhead{ID} &
    \colhead{$\log M_\star$} &
    \colhead{$\beta_\mathrm{UV, \texttt{CIGALE}}$} &
    \colhead{$\beta_\mathrm{UV, UV}$} &
    \colhead{${E(\mathit{B}-\mathit{V})}_{\mathrm{star, \texttt{CIGALE}}}$} &
    \colhead{${E(\mathit{B}-\mathit{V})}_{\mathrm{star, UV}}$} &
    \colhead{$\log \mathrm{SFR}_\mathtt{CIGALE}$} &
    \colhead{$\log \mathrm{SFR}_\mathrm{UV}$} &
    \colhead{$\log \mathrm{Age}_\mathtt{CIGALE}$} \\
    \colhead{} &
    \colhead{($M_\odot$)} &
    \colhead{} &
    \colhead{} &
    \colhead{(mag)} &
    \colhead{(mag)} &
    \colhead{($M_\odot{\,}\text{yr}^{-1}$)} &
    \colhead{($M_\odot{\,}\text{yr}^{-1}$)} &
    \colhead{($\text{yr}$)} \\
    \colhead{(1)} &
    \colhead{(2)} &
    \colhead{(3)} &
    \colhead{(4)} &
    \colhead{(5)} &
    \colhead{(6)} &
    \colhead{(7)} &
    \colhead{(8)} &
    \colhead{(9)}
  }
  \rotate{}
  \startdata{}
  312786 & $ \phn{}9.58 \pm 0.15 $ & $ -1.78 \pm 0.17 $ & $ -1.84 \pm 0.43 $ & $ 0.06 \pm 0.02 $ & $ 0.07 \pm 0.04 $ & $ 1.22 \pm 0.26 $ & $ 1.24 \pm 0.09 $ & $ 8.79 \pm 0.27 $ \\
  314763 & $ \phn{}9.55 \pm 0.20 $ & $ -1.17 \pm 0.14 $ & $ -1.24 \pm 0.22 $ & $ 0.14 \pm 0.02 $ & $ 0.12 \pm 0.02 $ & $ 1.95 \pm 0.25 $ & $ 1.70 \pm 0.05 $ & $ 8.14 \pm 0.39 $ \\
  324921 & $ \phn{}9.83 \pm 0.14 $ & $ -1.28 \pm 0.15 $ & $ -1.41 \pm 0.25 $ & $ 0.11 \pm 0.02 $ & $ 0.11 \pm 0.02 $ & $ 1.66 \pm 0.18 $ & $ 1.60 \pm 0.06 $ & $ 8.60 \pm 0.27 $ \\
  337592 & $ \phn{}8.29 \pm 0.19 $ & $ -2.13 \pm 0.13 $ & $ -2.16 \pm 0.54 $ & $ 0.05 \pm 0.02 $ & $ 0.04 \pm 0.05 $ & $ 1.41 \pm 0.15 $ & $ 0.96 \pm 0.12 $ & $ 7.34 \pm 0.64 $ \\
  340544 & $ \phn{}8.95 \pm 0.31 $ & $ -1.89 \pm 0.13 $ & $ -1.77 \pm 0.33 $ & $ 0.07 \pm 0.02 $ & $ 0.07 \pm 0.03 $ & $ 1.74 \pm 0.22 $ & $ 1.38 \pm 0.07 $ & $ 7.88 \pm 0.72 $ \\
  350488 & $ 10.36 \pm 0.08 $ & $ -0.20 \pm 0.37 $ & $ -1.16 \pm 0.73 $ & $ 0.18 \pm 0.05 $ & $ 0.13 \pm 0.06 $ & $ 1.23 \pm 0.24 $ & $ 1.10 \pm 0.16 $ & $ 9.12 \pm 0.16 $ \\
  796059 & $ \phn{}9.99 \pm 0.10 $ & $ -1.64 \pm 0.13 $ & $ -1.89 \pm 0.26 $ & $ 0.07 \pm 0.02 $ & $ 0.06 \pm 0.02 $ & $ 1.36 \pm 0.10 $ & $ 1.39 \pm 0.06 $ & $ 8.97 \pm 0.20 $ \\
  797658 & $ \phn{}9.02 \pm 0.24 $ & $ -1.46 \pm 0.19 $ & $ -1.56 \pm 0.51 $ & $ 0.11 \pm 0.03 $ & $ 0.09 \pm 0.05 $ & $ 1.47 \pm 0.31 $ & $ 1.17 \pm 0.11 $ & $ 8.20 \pm 0.52 $ \\
  798507 & $ \phn{}8.88 \pm 0.14 $ & $ -1.71 \pm 0.10 $ & $ -1.58 \pm 0.25 $ & $ 0.10 \pm 0.01 $ & $ 0.09 \pm 0.02 $ & $ 1.90 \pm 0.15 $ & $ 1.53 \pm 0.06 $ & $ 7.43 \pm 0.43 $ \\
  809877 & $ \phn{}8.46 \pm 0.07 $ & $ -2.26 \pm 0.07 $ & $ -2.31 \pm 0.27 $ & $ 0.04 \pm 0.01 $ & $ 0.03 \pm 0.02 $ & $ 1.66 \pm 0.09 $ & $ 1.19 \pm 0.06 $ & $ 7.15 \pm 0.24 $ \\
  810838 & $ \phn{}9.91 \pm 0.13 $ & $ -1.37 \pm 0.18 $ & $ -1.52 \pm 0.35 $ & $ 0.10 \pm 0.02 $ & $ 0.10 \pm 0.03 $ & $ 1.45 \pm 0.16 $ & $ 1.46 \pm 0.08 $ & $ 8.84 \pm 0.23 $ \\
  818357 & $ \phn{}8.75 \pm 0.32 $ & $ -1.30 \pm 0.26 $ & $ -1.40 \pm 0.85 $ & $ 0.13 \pm 0.03 $ & $ 0.11 \pm 0.08 $ & $ 1.56 \pm 0.25 $ & $ 1.07 \pm 0.19 $ & $ 7.98 \pm 0.87 $ \\
  870659 & $ \phn{}8.53 \pm 0.25 $ & $ -1.66 \pm 0.17 $ & $ -2.01 \pm 0.76 $ & $ 0.10 \pm 0.02 $ & $ 0.05 \pm 0.07 $ & $ 1.54 \pm 0.19 $ & $ 0.94 \pm 0.16 $ & $ 7.61 \pm 0.92 $ \\
  871455 & $ \phn{}8.87 \pm 0.26 $ & $ -2.14 \pm 0.11 $ & $ -2.06 \pm 0.30 $ & $ 0.05 \pm 0.02 $ & $ 0.05 \pm 0.03 $ & $ 1.55 \pm 0.21 $ & $ 1.27 \pm 0.07 $ & $ 7.91 \pm 0.54 $ \\
  891150 & $ \phn{}9.82 \pm 0.16 $ & $ -1.39 \pm 0.14 $ & $ -1.34 \pm 0.21 $ & $ 0.10 \pm 0.02 $ & $ 0.11 \pm 0.02 $ & $ 1.79 \pm 0.18 $ & $ 1.77 \pm 0.05 $ & $ 8.48 \pm 0.29 $ \\
  893160 & $ \phn{}8.30 \pm 0.17 $ & $ -2.27 \pm 0.10 $ & $ -2.16 \pm 0.48 $ & $ 0.04 \pm 0.01 $ & $ 0.04 \pm 0.04 $ & $ 1.32 \pm 0.16 $ & $ 1.00 \pm 0.10 $ & $ 7.45 \pm 0.49 $ \\
  894641 & $ \phn{}9.03 \pm 0.27 $ & $ -1.86 \pm 0.23 $ & $ -1.24 \pm 0.76 $ & $ 0.06 \pm 0.03 $ & $ 0.12 \pm 0.07 $ & $ 1.20 \pm 0.39 $ & $ 1.23 \pm 0.16 $ & $ 8.55 \pm 0.46 $ \\
  919252 & $ \phn{}8.72 \pm 0.15 $ & $ -2.06 \pm 0.10 $ & $ -2.17 \pm 0.27 $ & $ 0.06 \pm 0.01 $ & $ 0.04 \pm 0.02 $ & $ 1.75 \pm 0.15 $ & $ 1.30 \pm 0.06 $ & $ 7.42 \pm 0.46 $ \\
  919704 & $ 10.57 \pm 0.06 $ & $ -1.39 \pm 0.18 $ & $ -1.03 \pm 0.57 $ & $ 0.04 \pm 0.03 $ & $ 0.14 \pm 0.05 $ & $ 0.79 \pm 0.21 $ & $ 1.32 \pm 0.13 $ & $ 9.16 \pm 0.11 $ \\
  \tableline{}
  Low-mass composite & $ \phn{}8.74 \pm 0.33 $ & $-1.89 \pm 0.37$ & $-2.01 \pm 0.35$ & $0.06 \pm 0.03$ & $0.05 \pm 0.03$ & $1.55 \pm 0.21$ & $1.19 \pm 0.18$ & $7.61 \pm  0.42$\\
  High-mass composite & $ \phn{}9.87 \pm 0.31$ & $-1.38 \pm 0.23$ & $-1.37 \pm 0.27$ & $0.10 \pm 0.05$ & $0.11 \pm 0.02$ & $1.40 \pm 0.32$ & $1.42 \pm 0.27$ & $8.82 \pm 0.31$
  \enddata{}
  \tablecomments{
    (1) Object ID{};
    (2) stellar mass from SED fitting;
    (3) UV $\beta$ slope from the best-fit SED{};
    (4) UV $\beta$ slope from broad-band photometry;
    (5) $E(\mathit{B}-\mathit{V})$ for stellar continuum from SED fitting;
    (6) $E(\mathit{B}-\mathit{V})$ for stellar continuum from UV $\beta$ slope;
    (7) SFR from the best-fit SED{};
    (8) attenuation corrected SFR based on UV luminosity; and
    (9) stellar population age from the best-fit SED{}.
  }
\end{deluxetable*}

\subsection{UV-based dust attenuation and star formation rate}\label{sec:uvprop}

Measurements of the amount of dust attenuation and SFR are carried out by using UV part of SED,
namely the UV spectral slope \uvbeta defined as $f_\lambda \propto \lambda^{\uvbeta}$
and attenuation corrected far-UV (FUV) luminosity, respectively.
We derived \uvbeta with the identical way as \citet{onodera:2016}.
Briefly, we fit the total broad-band photometry in $r$, $ip$, $zpp$, $Y$, and $J$ bands
with a linear function in the magnitude--$\log\lambda$ space
to obtain far-UV (FUV{}; $\lambda\simeq1530$\AA) and near-UV (NUV{}; $\lambda\simeq2300$\AA) magnitudes
to compute \uvbeta following the prescription presented by \citet{pannella:2015}.
In the fitting, we put an additional weight of
$1 + \left | \lambda - \lambda_\text{FUV} \right | / \lambda_\text{FUV}$
in order to give more weights to filters closer to the rest-frame far-UV band \citep{nordon:2013}.
This procedure was repeated 5000 times by artificially perturbing the photometry with the corresponding errors.
The median values and standard deviations based on median absolute deviation ($\sigma_\text{MAD}$)
of the resulting \uvbeta and rest-frame FUV magnitude distributions are
considered as the values and associated $1\sigma$ errors, respectively.

The \uvbeta was then transformed into \ebvstar by using a recent calibration
by \citet{reddy:2018} for the SMC extinction curve \citep{gordon:2003},
\begin{equation}
    \uvbeta = -2.616 + 11.259\, \ebvstar.
    \label{eq:ebv_r18_smc}
\end{equation}
The intrinsic slope of $\uvbeta=-2.616$ is derived
by assuming a simple stellar population
of the Binary Population and Spectral Synthesis (BPASS) models \citep{eldridge:2017}
with a constant star formation with an age of 100~Myr and metallicity of $Z=0.002$,
with a two-power-law IMF with $\alpha=2.35$ for $M_\star>0.5\,M_\odot$
and $\alpha=1.30$ for $0.1 < M_\star/M_\odot < 0.5$.
The adopted relationship in \Cref{eq:ebv_r18_smc} provides
\ebvstar closer to that of the original calibration by \citet{calzetti:2000}
with discrepancies within $\simeq 0.1$~mag, 
while using the relations derived by the same authors \citep{reddy:2015}
for the \citet{calzetti:2000} and \citet{reddy:2015} attenuation curves
provides systematically higher \ebvstar than that from the SMC curve
by $0.03$--$0.2$~mag.

By using the derived \ebvstar, SFRs were computed from attenuation corrected
rest-frame FUV (1500~\AA) luminosities ($L_{1500}$) using the calibration by \citet{theios:2019},
\begin{equation}
    \log \mathrm{SFR} / (M_\odot{\,}\text{yr}^{-1}) = \log \nu L_\nu (1500\,\text{\AA}) / (\text{erg}\,\text{s}^{-1}) - 43.46.
    \label{eq:lum2sfr}
\end{equation}
This relation assumes the same intrinsic stellar population model as used to derive \ebvstar from \uvbeta,
and is slightly different from that of \citet{kennicutt:2012} where the constant term is 43.35.

\begin{figure*}
    \centering
    \includegraphics[width=0.8\linewidth]{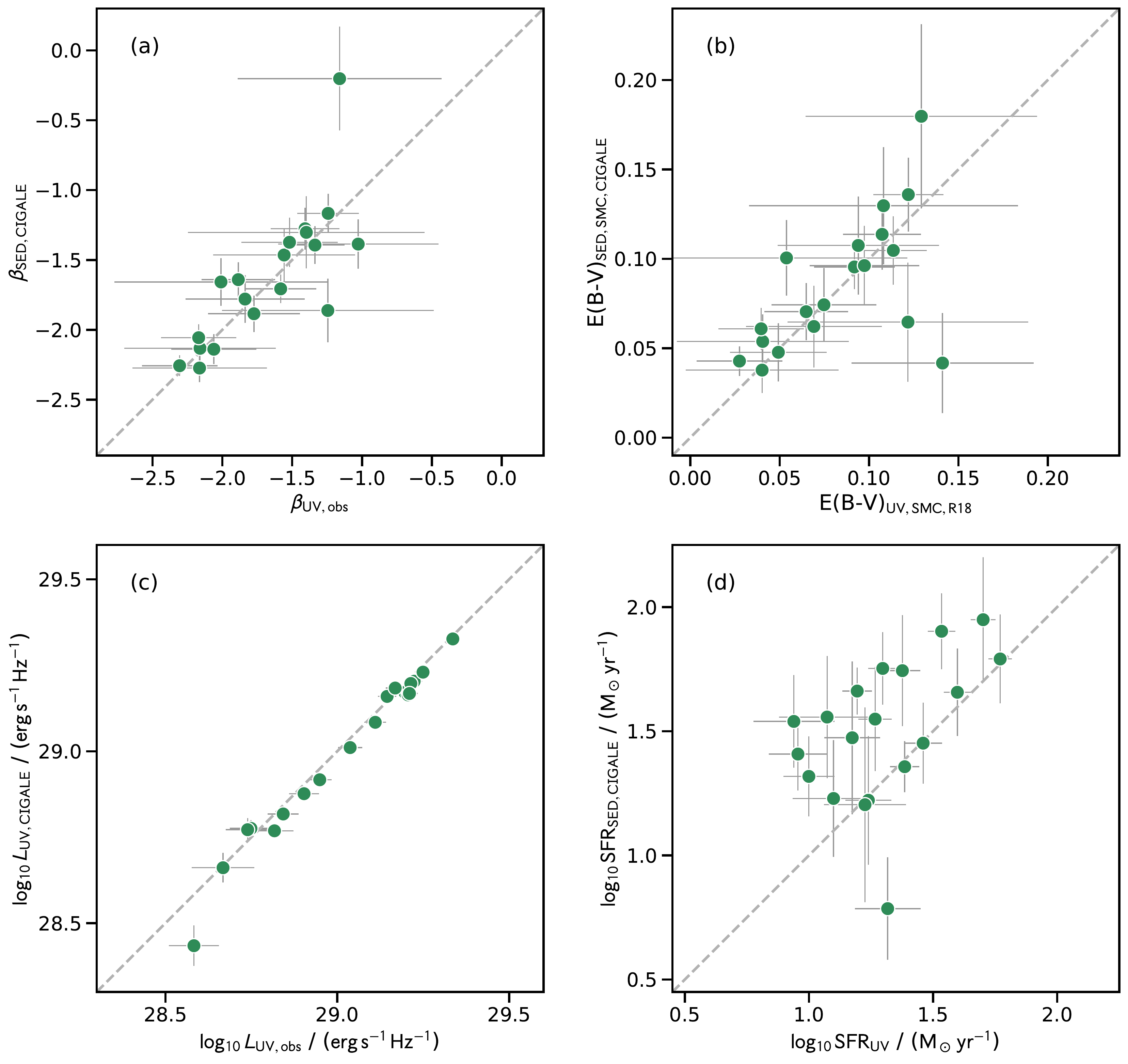}
    \caption{
        Comparison of UV SED properties and SFR derived based on observed UV photometry (horizontal axes) and
        on SED fitting using \cigale (vertical axes).
        (a) UV $\beta$ slope;
        (b) stellar $E(\mathit{B}-\mathit{V})$ assuming SMC extinction curve;
        (c) UV luminosity at rest-frame 1500~\AA{} without extinction correction; and
        (d) extinction-corrected SFR{}.
        \label{fig:comp_dust}
    }
\end{figure*}

The attenuation properties and SFRs derived here are shown in \Cref{tab:sedprop}.
These properties are compared with those derived from SED fitting in \Cref{fig:comp_dust}.
Both estimates agree well for \uvbeta, \ebvstar,
and $L_\nu(1500\,\mathrm{\AA}{})$ with a few outliers which are caused mainly
due to large uncertainties in the rest-frame UV photometric data.
SFRs derived by \cigale SED fitting is systematically higher than those
based on UV luminosities with a median difference of $0.28$~dex.
The systematic difference is likely due to different assumptions in the intrinsic SED and dust attenuation.
In the SED fitting, we used \citet{bruzual:2003} models,
while BPASS models \citep{eldridge:2017} are used in the calibration by \citet{reddy:2018} in the UV-based estimate.
Moreover, the relation between \ebvstar and \ebvneb applied for the \cigale run is derived
assuming BPASS models and a revised version of SMC extinction curve \citep{reddy:2016:dust,theios:2019}.
Reconciling the two estimates is not the scope of this paper, and
we will use only stellar masses from the SED fitting
and use UV-based estimates for the attenuation and SFR parameters in the following analysis.

\subsection{Equivalent width of the \oiiitwo emission line}\label{sec:eq}

\begin{deluxetable*}{ccccc}

  \tablecaption{
    Ionization properties
    \label{tab:ismprop}
  }
  \tablehead{
    \colhead{ID} &
    \colhead{$\text{EW} (\oiiitwo)$} &
    \colhead{$O_{32}$} &
    \colhead{$R_{23}$} &
    \colhead{$\log \xiionz$}\\
    \colhead{} &
    \colhead{(\AA)} &
    \colhead{} &
    \colhead{} &
    \colhead{($\mathrm{erg}^{-1}\,\mathrm{Hz}$)}\\
    \colhead{(1)} &
    \colhead{(2)} &
    \colhead{(3)} &
    \colhead{(4)} &
    \colhead{(5)}
  }
  \startdata{}
  312786 & $\phn{}223 \pm 11$ & $>4.5$ & $>10.5$ & $<25.0$ \\
  314763 & $\phn{}217 \pm \phn{}9$ & $1.9 \pm 0.7$ & $\phn{}6.5 \pm 1.7$ & $25.30 \pm 0.07$ \\
  324921 & $\phn{}250 \pm \phn{}9$ & $1.5 \pm 0.6$ & $\phn{}9.6 \pm 2.8$ & $25.33 \pm 0.08$ \\
  337592 & $2051 \pm 82$ & $>7.2$ & $\phn{}4.8 \pm 3.1$ & $25.86 \pm 0.15$ \\
  340544 & $1391 \pm 57$ & $2.9 \pm 1.3$ & $\phn{}9.4 \pm 3.4$ & $25.63 \pm 0.09$ \\
  350488 & $\phn{}233 \pm 11$ & $>4.3$ & $>8.5$ & $<25.7$ \\
  796059 & $\phn{}407 \pm 16$ & $2.7 \pm 1.2$ & $>6.5$ & $<25.7$ \\
  797658 & $1123 \pm 33$ & \nodata{} & \nodata{} & $<26.0$ \\
  798507 & $\phn{}557 \pm 30$ & $2.5 \pm 1.0$ & $\phn{}8.1 \pm 2.8$ & $25.30 \pm 0.11$ \\
  809877 & $1853 \pm 64$ & \nodata{} & \nodata{} & $<25.8$ \\
  810838 & $\phn{}262 \pm 14$ & $1.5 \pm 0.7$ & $\phn{}5.2 \pm 2.0$ & $25.70 \pm 0.11$ \\
  818357 & $\phn{}670 \pm 60$ & $>3.0$ & $>4.6$ & $<25.7$ \\
  870659 & $\phn{}570 \pm 43$ & $>4.2$ & $\phn{}8.8 \pm 8.1$ & $25.21 \pm 0.23$ \\
  871455 & $\phn{}472 \pm 29$ & $>5.8$ & $>9.0$ & $<25.0$ \\
  891150 & $\phn{}153 \pm 13$ & $1.3 \pm 0.4$ & $>11.2$ & $<25.0$ \\
  893160 & $1171 \pm 58$ & $>5.3$ & $\phn{}5.0 \pm 2.8$ & $25.54 \pm 0.13$ \\
  894641 & $\phn{}572 \pm 22$ & $>2.6$ & $\phn{}7.7 \pm 6.8$ & $25.32 \pm 0.20$ \\
  919252 & $\phn{}860 \pm 27$ & $3.3 \pm 1.3$ & $14.7 \pm 5.0$ & $25.15 \pm 0.10$ \\
  919704 & $\phn{}133 \pm \phn{}5$ & $>2.5$ & $\phn{}5.7 \pm 3.8$ & $25.58 \pm 0.16$ \\
  \tableline{}
  Low-mass composite & $\phn{}791 \pm 360$ & $5.7 \pm 2.5$ & $10.8 \pm 3.7$ & $25.32 \pm 0.18$ \\
  High-mass composite & $\phn{}189 \pm \phn{}30$ & $2.7 \pm 0.5$ & $\phn{}9.6 \pm 1.5$ & $25.43 \pm 0.27$
  \enddata{}
  \tablecomments{
    (1) Object ID{};
    (2) rest-frame equivalent width of \oiiitwo;
    (3) $O_{32}$ as defined in \Cref{eq:o32};
    (4) $R_{23}$ as defined in \Cref{eq:r23}; and
    (5) \xiionz defined in \Cref{eq:xiion} assuming $\fesc=0$.
  }
\end{deluxetable*}

Because our EELG sample does not show continuum in the MOIRCS spectra in general (\Cref{fig:spec2d}),
emission line equivalent widths cannot be measured directly on the spectra.
Furthermore, since \textit{Ks} broad-band fluxes are dominated by $\oiii + \hbeta$ up to $\simeq 100$~\%{},
estimated continuum levels based on the \textit{Ks}-band photometry can be comparable to the photometric error.
Therefore, we use the best-fit stellar SED derived in \autorefsec{sec:mass} assuming no error in the continuum.
The model continuum is derived as the median flux at $4940$--$5028$~\AA{} in the rest-frame wavelength.
The derived $\text{EW} (\oiiitwo)$ is shown in \Cref{tab:ismprop}.
It should be noted that the assumption of the error-free continuum is fairly strong.
If we assume the same errors as the total \textit{Ks}-band fluxes for the continuum,
the resulting uncertainties of $\text{EW} (\oiiitwo)$ increase from $\simeq 0.01$--$0.02$~dex to $\simeq 0.04$~dex,
which does not change any of our results and conclusions presented in this study.

\begin{figure}
    \centering
    \includegraphics[width=0.95\linewidth]{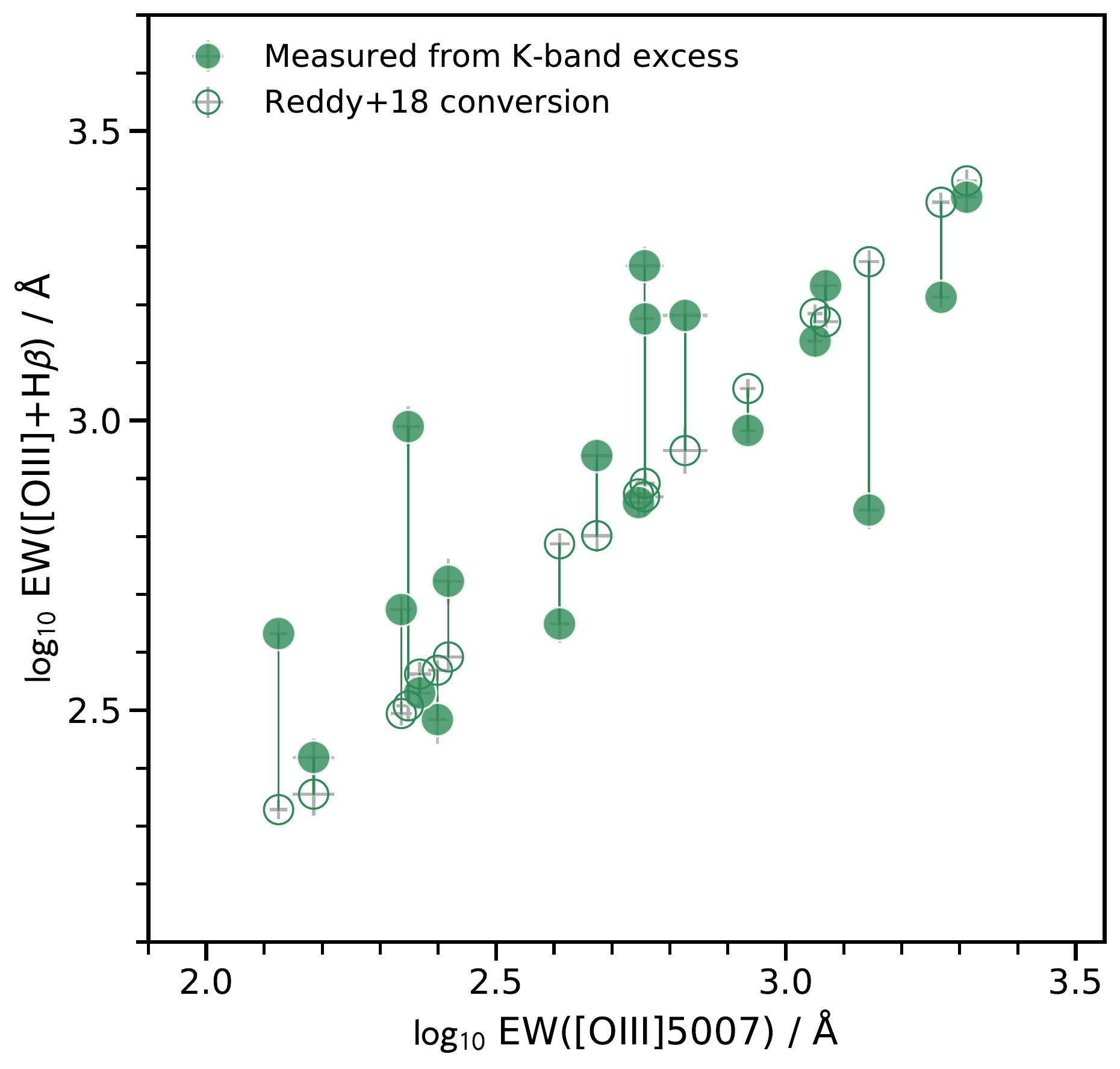}
    \caption{
        Comparison of $\text{EW}(\oiiitot + \hbeta)$ derived with two different methods
        with the spectroscopically measured $\text{EW}(\oiiitwo)$.
        Filled circles show $\text{EW}(\oiiitot + \hbeta)$ derived from the \textit{Ks}-band photometric excess, \deltak,
        while open circles are predictions using a correction factor derived by \citet{reddy:2018:ew}.
        Vertical lines connect each object.
        \label{fig:comp_o3ew}
    }
\end{figure}

In \Cref{fig:comp_o3ew}, we compare $\text{EW} (\oiiitwo)$ with $\text{EW} (\oiii + \hbeta)$ estimated from \deltak and that predicted with a conversion derived by \citet{reddy:2018:ew}
for a sample of star-forming galaxies at $z \simeq 3.4$.
The former is derived by using \Cref{eq:deltakew},
while the latter is estimated as
\begin{equation}
    \text{EW} (\oiii + \hbeta) = c \, \text{EW} (\oiii).
\end{equation}
The correction factor $c$ is defined as $c = 0.581 - 0.0074x + 6.322\times 10^{-3}x^2$ where $x=\log M_\star/M_\odot$ \citep{reddy:2018:ew}.
Although about a half of our sample has lower stellar mass
than the range of $9.0 < \log M / M_\odot < 11.5$ for which the conversion is derived,
difference between two estimates agrees well
with a median of 0.06 dex and a standard deviation of 0.13 dex
which is smaller than the $\simeq 0.2$--$0.3$~dex scatter
of the sample distribution presented in \citet{reddy:2018:ew}.

\subsection{$R_{23}$ and $O_{32}$ line ratios\label{sec:r23o32}}

We derived $R_{23}$ and $O_{32}$ indices defined as
\begin{equation}
    R_{23} \equiv \frac{\oii + \oiiitot}{\hbeta}
    \label{eq:r23}
\end{equation}
and
\begin{equation}
    O_{32} \equiv \frac{\oiiitot}{\oii},
    \label{eq:o32}
\end{equation}
respectively.
All emission line fluxes are corrected for
dust extinction using \uvbeta-based \ebvneb (see \autorefsec{sec:uvprop})
with a conversion $\ebvneb = 3.06\, \ebvstar$
for the average SMC Bar extinction curve \citep{gordon:2003,theios:2019}.

We assumed $\oiiitwo / \oiiione = 3$ to estimate the total [\ion{O}{3}] flux.
When \oii emission lines are not detected, we assign fluxes corresponding to the $3\sigma$ upper limit
to derive the lower limit of $O_{32}$, while undetected \oii fluxes are assumed to be zero for $R_{23}$
because they are essentially negligible compared to those of \oiiitwo.
These upper limits are also corrected for dust extinction as described above.

Detected \hbeta fluxes are also corrected for stellar absorption by using the best-fit SED derived in \autorefsec{sec:mass}.
The correction factors are $\lesssim 10$~\%{} for less massive galaxies with $\log M_\star/M_\odot < 10$,
while one object with $\log M_\star/M_\odot\simeq 10.5$ has a correction factor of $\simeq 40$~\%{}.
The latter value is slightly larger than the prediction using the correction factor--stellar mass relation
derived by \citet{zahid:2014} for a sample of star-forming galaxies at $z \simeq 1.6$.

The derived $O_{32}$ and $R_{23}$ values are listed in \Cref{tab:ismprop}.

\subsection{Ionizing photon production efficiency $\xiion$}

The LyC photon production efficiency, $\xiion$, is defined as
the ratio of the production rate of LyC photons, $N(\mathrm{H}^{0})$ to
the attenuation-corrected UV continuum luminosity density, $L_\mathrm{UV,corr}$:
\begin{equation}
    \xiion = \frac{N (\mathrm{H}^{0})}{L_\mathrm{UV,corr}}\,\left[\mathrm{s}^{-1} / (\mathrm{erg}\,\mathrm{s}^{-1}\,\mathrm{Hz}^{-1})\right].
    \label{eq:xiion}
\end{equation}
The UV luminosity is derived at the rest-frame 1500~\AA{}
using broad-band photometry (\autorefsec{sec:uvprop}).
For $N(\mathrm{H}^{0})$, we adopt the relation to the intrinsic \halpha luminosity, $L(\halpha)$,
by \citet{leitherer:1995},
\begin{equation}
    L (\halpha)\,\left[\mathrm{erg}\,\mathrm{s}^{-1}\right] =
    1.36 \times 10^{-12}\,N (\mathrm{H}^{0})\,\left[\mathrm{s}^{-1}\right].
    \label{eq:nh}
\end{equation}
Then we assume $L(\halpha)/L(\hbeta) = 2.86$ assuming the case B recombination
with the electron density of 100 $\mathrm{cm}^{-3}$ and temperature of $10^4\,\mathrm{K}$
\citep{osterbrock:agnagn} to compute \xiion for our EELG sample at $z \simeq 3.3$.
The \hbeta fluxes are corrected for dust attenuation and stellar absorption
as described in \autorefsec{sec:r23o32}.
One should keep in mind that the relation above is derived
for an ionization-bounded \ion{H}{2} region with no LyC escape.

LyC escape fraction is suggested to be correlated with some of galaxy properties
such as $\mathit{O}_{32}$ \citep[e.g.,][]{nakajima:2014,nakajima:2020:laces,faisst:2016:lyc,izotov:2018:lyc},
\ebv \citep[e.g.,][]{reddy:2016:ionize},
and \lya profile \citep[e.g.,][]{dijkstra:2016,verhamme:2015,izotov:2020:lya}.
Although we have measurements of $\mathit{O}_{32}$ and \ebv in hand,
we do not correct $\fesc$ here,
because the correlations between \fesc and these parameters show large scatters
especially at large $\mathit{O}_{32}$ and low \ebv
\citep[e.g.,][]{faisst:2016:lyc,reddy:2016:ionize,nakajima:2020:laces}.
Instead, we use \xiionz for the case of zero LyC escape fraction \citep{bouwens:2016,nakajima:2016}
to explicitly indicate the assumption of no LyC escape from the \ion{H}{2} region in \Cref{eq:nh}.
The intrinsic \xiion then can be derived by $\xiion = \xiionz / (1 - \fesc)$.
We will investigate the relationship between \xiionz
and other galaxy properties in \autorefsec{sec:xiion}.
The derived \xiionz are shown in \Cref{tab:ismprop}.

\subsection{Composite spectra}\label{sec:stack}

\begin{figure*}
    \centering
    \includegraphics[width=0.95\linewidth]{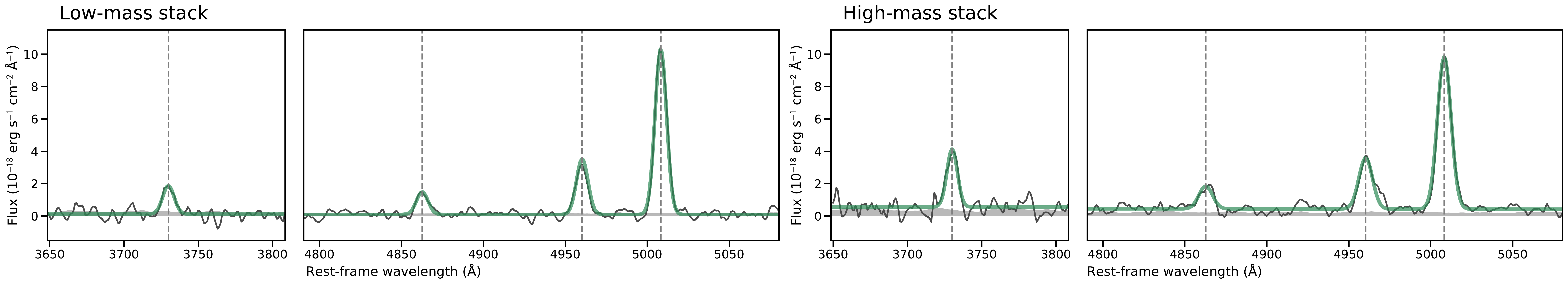}
    \caption{
        Stacked spectra of our EELGs at $z \simeq 3.3$ for
        low-mass ($\log M_\star/M_\odot \lesssim 9$) bin (two leftmost panels)
        and high-mass ($\log M_\star/M_\odot \gtrsim 9.5$) bin (two rightmost panels).
        Contents of each panel are the same as those in \Cref{fig:spec1d}
        except that the horizontal axis shows the rest-frame wavelength.
        \label{fig:stackspec}
    }
\end{figure*}

In order to investigate average spectroscopic properties of EELGs at $z\simeq 3.3$,
we create composite Subaru/MOIRCS spectra.
We split the sample in the stellar mass at $\log M_\star/M_\odot \simeq 9.2$.
There are 11 and 8 objects in the low-mass and high-mass bins, respectively.
Before the stacking, each 1D spectrum is shifted and linearly interpolated
to the rest-frame wavelength grid of 1~\AA{} interval,
while the corresponding noise spectrum is interpolated to the same wavelength grid in quadrature.
Then composite spectra are constructed by taking an average
at each wavelength weighted by the inverse variance.
The associated noise spectra are derived by the standard error propagation
from the individual noise spectra.
Emission line fluxes are measured by fitting Gaussian components for emission lines
and linear continuum similar to that applied to individual objects (\autorefsec{sec:specmeas}).
The resulting stacked spectra are shown in \Cref{fig:stackspec}
and emission line fluxes are shown in \Cref{tab:emfit}.

SED properties for the composite spectra are derived
by taking the median and $\sigma_\text{MAD}$ of the sample in each bin.
Equivalent widths of \oiiitwo, emission line ratios, and \xiionz
of the composite spectra are derived
by using the derived SED properties as described above
and emission line fluxes measured on the composite spectra.
These parameters are appended in \Cref{tab:sedprop} and \Cref{tab:ismprop}.

\section{Results and discussion} \label{sec:results}

\subsection{Testing COSMOS2015 photometric redshifts and SED fitting}\label{sec:compz}

The COSMOS2015 catalog provides photometric redshifts
with a precision of $\sigma_{\Delta{z/(1+z)}}=0.021$ with a catastrophic failure of $13.2$~\%{}
for star-forming galaxies at $z>3$ \citep{laigle:2016}.
Although nebular emission line contributions of \lya, \oii, \hbeta, \oiiitot, and \halpha
have already taken into account when deriving the photometric redshifts,
their emission line ratios are fixed to representative values
for normal star-forming galaxies.
For example, while they fixed $O_{32}=0.36$,
our EELG sample at $z\simeq 3.3$ shows a significantly higher $O_{32}$ with the median of $\gtrsim 3$.
Therefore, it is worth checking whether the photometric redshift and physical parameters
in the COSMOS2015 catalog work for our EELG sample or not.

\begin{figure*}
    \centering
    \includegraphics[width=0.99\linewidth]{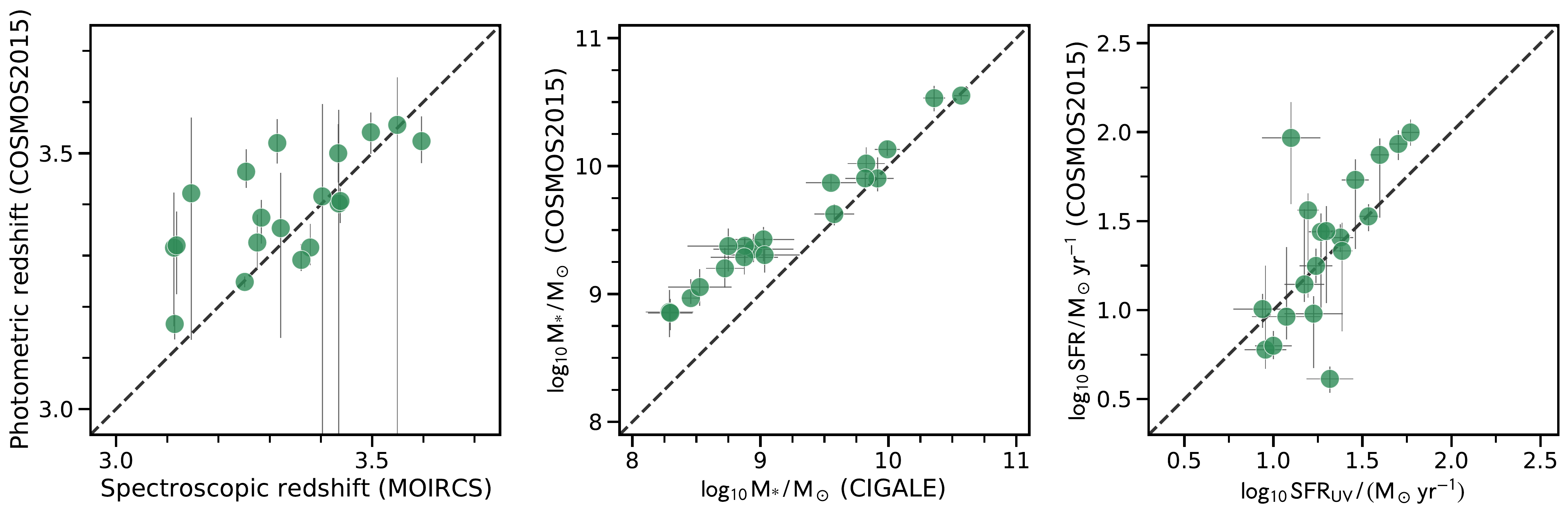}
    \caption{
        Comparison of photometric redshift (left), stellar mass (middle),
        and star formation rate (right)
        between the publicly available COSMOS2015 catalog \citep{laigle:2016}
        and this study.
        Dashed lines indicate 1-to-1 relation for each parameter.
        Only EELGs at $3<z<3.7$ in our sample are shown.
        \label{fig:comp_cosmos2015}
    }
\end{figure*}

In the left panel of \Cref{fig:comp_cosmos2015}, we compare the COSMOS2015 photometric redshifts
with spectroscopic ones for our identified EELG sample at $z>3$.
Using the same metric as \citet{laigle:2016},
we compute $\sigma_{\Delta_{z/(1+z)}}=0.025$ with a systematic offset of $-0.01$
and no catastrophic failure.
Two of our spectroscopic sample are at $z_\text{phot} \simeq 2.2$ and
their spectroscopic redshifts show almost the exact agreement.
There are two objects without emission line detections.
Since their COSMOS2015 photometric redshifts are 2.476 and 3.774, respectively,
the primary emission lines (\halpha and \oiiitwo, respectively)
may fall outside of the spectral coverage of MOIRCS{}.
Here, we conclude that the photometric redshift in COSMOS2015 for extreme emission line galaxies at $z\simeq3.3$
is as precise as those for other galaxies at the similar redshift
despite the fact that the emission line ratios assumed in the COSMOS2015 catalog is
significantly different from our sample.
Note, however, that the spectroscopic follow-up was carried out for a sample with visually reliable SED shapes
and no attempt was done to search for EELGs at $z\simeq3.3$
with $z_\text{phot} \lesssim 2$ and $z_\text{phot} \gtrsim 4$.

Stellar masses are compared in the middle panel of \Cref{fig:comp_cosmos2015}.
COSMOS2015 stellar masses are systematically larger than those derived by our SED fitting,
especially evident at lower stellar masses of $\log M_\star/M_\odot \lesssim 9$.
We do not find any indications that
the difference between two stellar mass estimates
correlates with overestimates of photo-\textit{z}.
Rather, this can be understood by a realistic treatment of
the emission line strengths and ratios especially of \oiii
to the \textit{Ks} band fluxes in \cigale using the recipe of \citet{inoue:2011}
compared to empirical line ratios in the COSMOS2015 SED fitting.
Therefore, in order to estimate more appropriate stellar masses,
one may not be able to rely on the catalog values
for low-mass ($\log M_{\star,\mathrm{COSMOS2015}}/M_\odot \lesssim 9.5$) galaxies at $z \gtrsim 3$
and needs to perform a dedicated SED fitting depending on the sub-population of interest.

On the other hand, SFRs between UV-based and COSMOS2015 estimates agree well.
At high SFR of $\log \text{SFR} \gtrsim 1.5$,
our UV-based SFRs seems to be slightly smaller than those in the COSMOS2015 catalog.
As shown in \Cref{fig:comp_dust}, UV-based SFRs tend to be lower than those derived in the \cigale SED fitting.
Therefore, if $\text{SFR}_\text{UV}$ is replaced to $\text{SFR}_\text{\cigale}$ in \Cref{fig:comp_cosmos2015},
points will move toward right and $\text{SFR}_\text{\cigale}$ would be higher than $\text{SFR}_\mathrm{COSMOS2015}$
by $\simeq 0.3$~dex.

Note also that the SED fitting in the COSMOS2015 catalog and this study use
different assumptions on the set of intrinsic SED templates,
dust attenuation recipes,
and the explored ranges of age, metallicity, dust attenuation,
and ionization parameters, which could introduce systematics
for the estimate of physical parameters.

\subsection{AGN contamination}\label{sec:agn}

Intense \oiii emission with a large EW
and large $\oiii / \hbeta$ ratio
can originate from AGN
as well as star-forming galaxies
\citep[e.g.][]{baskin:2005,kewley:2006,caccianiga:2011}.
Since our primary motivation in this study is
to search for a population of EELGs at $z \gtrsim 3$
as an analog of star-forming galaxies in the EoR,
we would like to check whether AGN dominate the sample or not.

Among 240 objects in the parent sample,
142 objects have $3<z_\text{phot}<3.7$ and
none of them show counterparts in X-ray
in the \textit{Chandra} COSMOS Legacy Survey catalog
\citep{civano:2016,marchesi:2016}
and in radio in the public VLA data.

There are 4 EELG candidates showing \textit{Spitzer}/MIPS $24\,\mu \text{m}$
detections ranging from 70--240~$\mu\text{Jy}$.
The $24\,\micron$ fluxes roughly correspond to
$\text{SFR}\gtrsim 1000\,M_\odot{\,}\text{yr}^{-1}$
using the \citet{dale:2002} templates to convert
$24\,\micron$ fluxes to total IR luminosities
and the \citet{kennicutt:1998} relation to
convert total IR luminosities to SFRs
\citep[e.g., ][]{wuyts:2008, marchesini:2010}.
If the $24\,\micron$ fluxes are due to star formation,
they are well above the MS at $z \simeq 3.3$ \citep[e.g.,][]{speagle:2014}
and they would be detected in far-IR to submillimeter wavelength.
However, none of 140 EELG candidates at $3<z_\text{phot}<3.7$
are detected in 100--500~\micron{}
in the \textit{Herschel}/PACS PEP survey \citep{lutz:2011} and
\textit{Herschel}/SPIRE HerMes survey \citep{oliver:2012}.
Therefore, these MIPS $24\,\micron$ sources can be hosts of obscured AGN,
but they are a rare among the selected EELG candidates, only $\lesssim 5$~\%{}.
Although the discussion above does not rule out the presence of low-luminosity or less obscured AGN
undetected in $24\,\micron$, based on the available information for emission line properties,
it is unlikely that AGN dominate the EELG population at $z \simeq 3.3$ as we discuss below.

The emission line widths of \oiiitwo of identified EELGs at $z\simeq3.3$
are $\text{FWHM} \simeq 400$--$500\,\kms$,
which is actually narrower than the expected resolution
of $960\,\kms$ of the MOIRCS HK500 grism.
This could be due to the combination of better seeing size
than the slit width and intrinsically compact nature of the objects.
Since none of our EELG sample at $z\simeq 3.3$ shows broad \oiii line
exceeding $\text{FWHM} \simeq 1000\,\kms$ indicative of
AGN-driven outflow \citep[e.g.,][]{holt:2008,alexander:2010,genzel:2014b},
the line widths are rather consistent with star formation.

\begin{figure}
    \centering
    \includegraphics[width=0.99\linewidth]{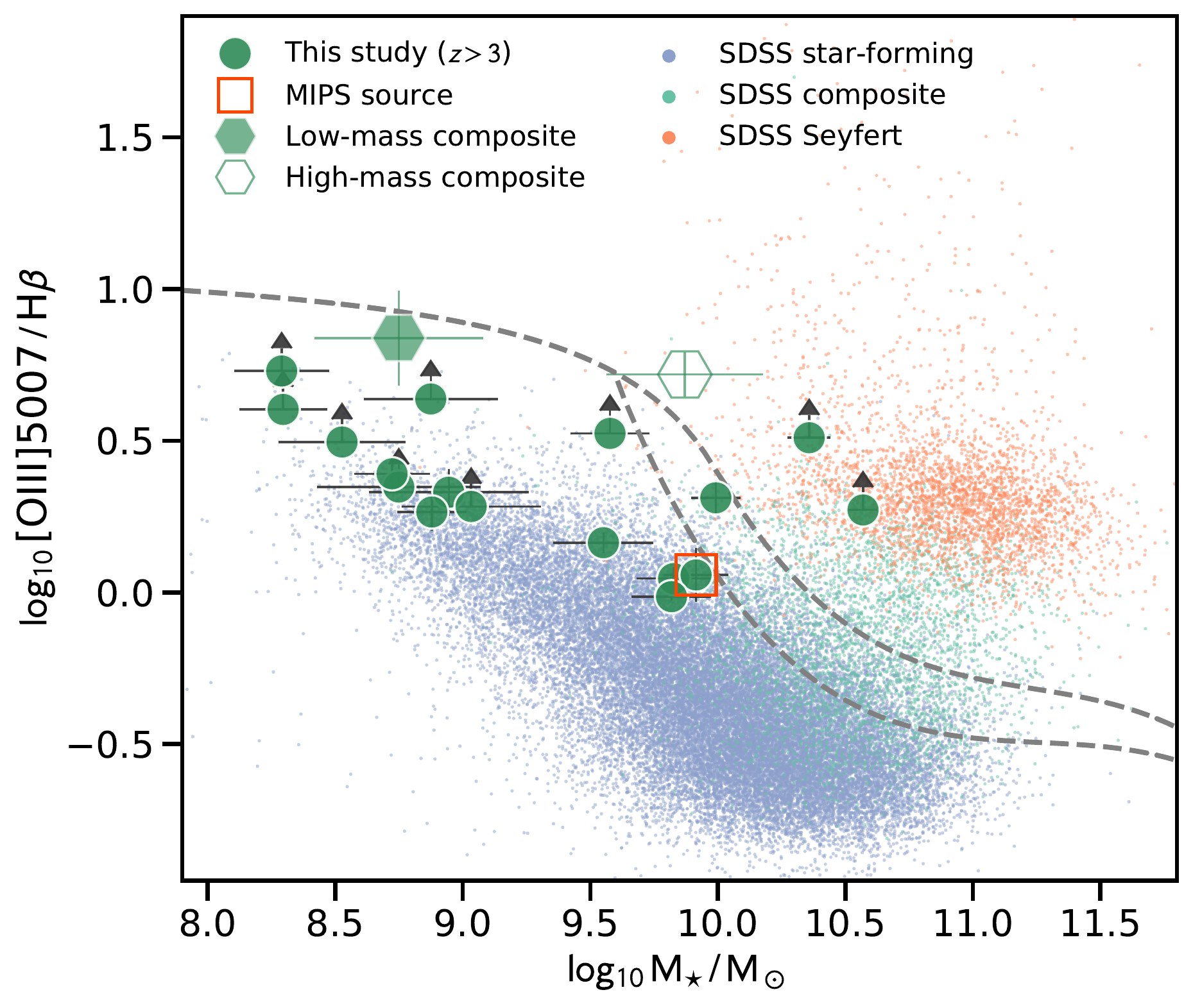}
    \caption{
        The MEx diagram for spectroscopically identified EELGs at $z \simeq 3.3$ (large circles).
        The red square indicates a MIPS 24~\micron{} source.
        Measurements on the low-mass and high-mass composite spectra
        are shown with filled and open hexagons, respectively.
        Small dots are galaxies from OSSY SDSS DR7 catalog \citep{oh:2011}
        classified as star-forming (blue),
        composite (green), and Seyfert (orange) galaxies
        using the BPT diagram \citep{kauffmann:2003, kewley:2006}.
        Dashed lines indicate the demarcation lines derived by \citep{juneau:2014}
        to separate star-forming, composite, and Seyfert galaxies.
        \label{fig:mex}
    }
\end{figure}

Emission line ratios can also be used to diagnose the ionization mechanism
through, e.g., BPT diagram \citep{baldwin:1981:bpt, kewley:2001, kauffmann:2003}.
However, for our EELGs at $z \simeq 3.3$, only \hbeta and \oiii lines are available.
As an alternative to the BPT diagram,
\citet{juneau:2011} proposed to use the Mass-Excitation (MEx) diagram
by using \oiiitwo/\hbeta and stellar mass.
\Cref{fig:mex} shows MEx diagram for our EELGs at $z\simeq 3.3$
and SDSS DR7 objects taken from the OSSY catalog \citep{oh:2011}
with the revised demarcation lines to classify star formation,
composite, and AGN \citep{juneau:2014}.  %
In the MEx diagram, most of the individual EELGs and low-mass composite are
in the region of either star-forming or composite galaxies,
but a couple of objects as well as the high-mass composite are classified as Seyfert galaxies.
These two objects are the most massive ones among the spectroscopically confirmed objects
and they show noticiably redder SEDs with old stellar populations of $\gtrsim 1$~Gyr
as seen in \Cref{fig:sedfitting} and \Cref{tab:sedprop}.
Moreover, they show significantly lower SFRs relative to
the MS of SFGs at $z \simeq 3.3$ (see \autorefsec{sec:mainsequence} and \Cref{fig:sfrmass}).
Therefore, they could be obscured AGN despite no indication in the other diagnostics.
On the other hand, one MIPS 24~\micron{} source is actually classified as a star-forming galaxy.
Note that the demarcation lines of \citet{juneau:2014} are, however, calibrated
by using local galaxies from SDSS{}.
\citet{strom:2017} showed that SFGs at $z\simeq 2.3$
from the KBSS-MOSFIRE survey distribute toward higher \oiii/\hbeta ratios
up to 0.8 dex at a given stellar mass in the MEx diagram.
Our sample classified as AGN by \citet{juneau:2014} lines
are indeed consistent with the KBSS-MOSFIRE star-forming sample at $z\simeq2.3$
as well as AGN when considering the lower limits more conservatively \citep{strom:2017}.

In summary, looking at various possible diagnostics of AGN for our sample,
we conclude that EELGs at $z\simeq 3.3$ are most likely to be dominated by star formation,
though we cannot completely rule out the presence of AGN at massive end ($M_\star \gtrsim 10^{10.5}\,M_\odot$).
In the following analysis, we consider the all spectroscopically identified
EELGs at $z \simeq 3.3$ as SFGs.

\subsection{Are EELGs at $z\simeq3.3$ on the main-sequence of star-forming galaxies?}\label{sec:mainsequence}

\begin{figure}
    \centering
    \includegraphics[width=0.99\linewidth]{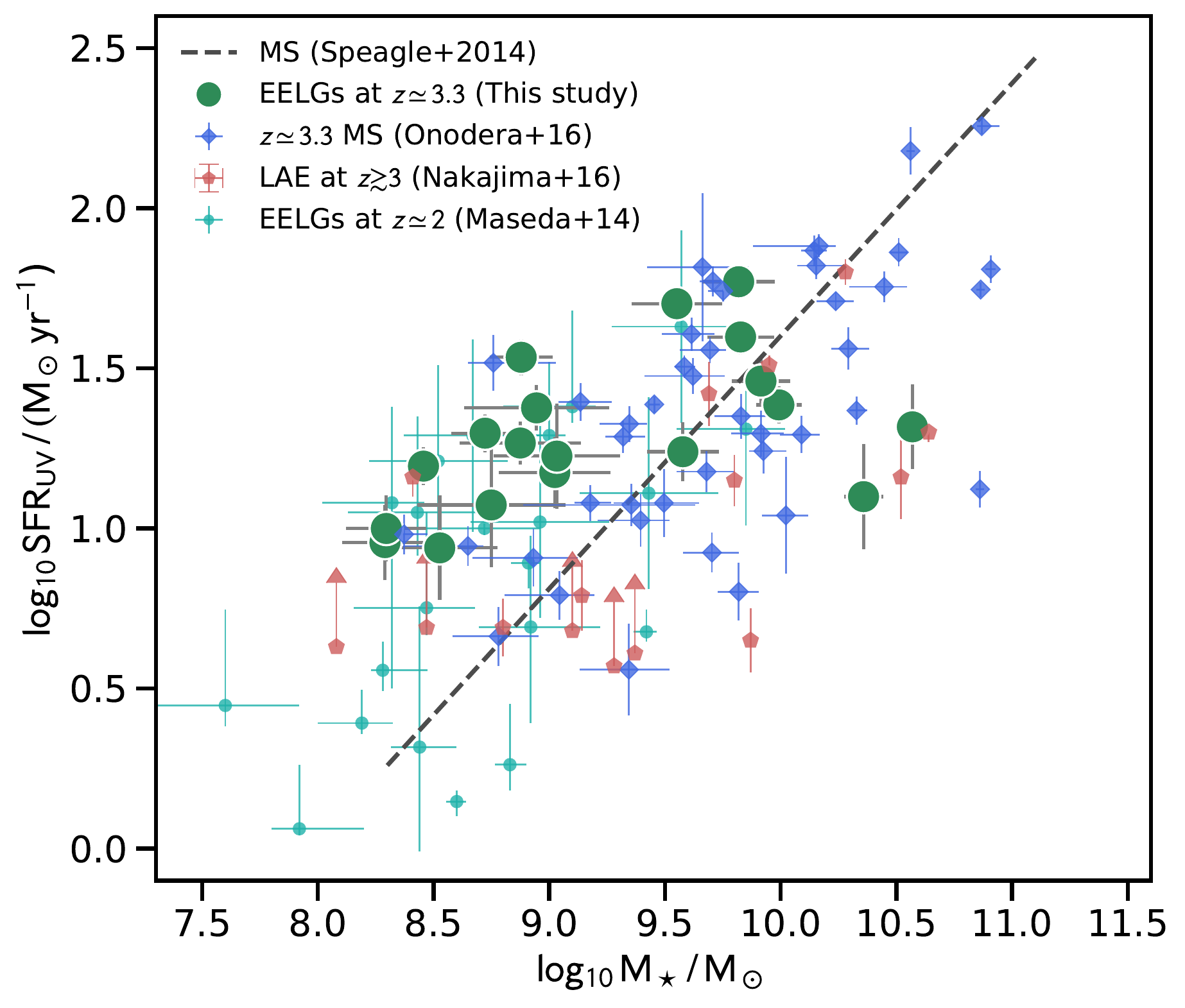}
    \caption{
        SFR as a function of stellar mass.
        Green circles show our EELG sample at $z \simeq 3.3$,
        small light green circles show EELGs at $1.3<z<2.3$ \citep{maseda:2014},
        blue diamonds show UV continuum selected SFGs at $z \simeq 3.3$ \citep{onodera:2016},
        and red diamonds show \lya emitters at $z=3$--$4$ \citep{nakajima:2016}.
        The dashed line indicates the MS at $z \simeq 3.3$ \citet{speagle:2014}.
        \label{fig:sfrmass}
    }
\end{figure}

\Cref{fig:sfrmass} shows the relation between SFR and stellar mass for
our EELG sample at $z\simeq 3.3$ together with normal star-forming galaxies at $z \simeq 3.3$
selected via UV-continuum brightness or expected \hbeta flux \citep{onodera:2016}
and \lya emitters (LAEs) at $z=3$--$4$ studied by \citet{nakajima:2016}.
Objects taken from \citet{nakajima:2016} and \citet{onodera:2016}
distribute around the fiducial MS of SFGs at $z \simeq 3.3$ \citep{speagle:2014}.
While our spectroscopically identified EELGs at $z \simeq 3.3$ are
on the MS at $\log M/M_\odot \gtrsim 9.5$,
those with $\log M/M_\odot \lesssim 9.0$ clearly show higher SFR than the MS by $\gtrsim 0.5$--$1$~dex.

The elevated SFR relative to the MS in less massive galaxies
is also seen in EELGs at $1.3<z<2.3$ identified by \citet{maseda:2014}
as also shown in \Cref{fig:sfrmass} with light green circles.
They conclude that the star formation in low mass galaxies
producing intense \oiii and \halpha emission lines
is undergoing a burst-like stage with a time scale of $\sim 50$~Myr.
The mass-weighted age from the SED fitting with \cigale for our EELG sample
are indeed young, $\simeq 10$--$100$~Myr, at $\log M/M_\odot \lesssim 9.0$.
Such young stellar population ages for low-mass galaxies
are also reported by \citet{vanderwel:2011} for a sample of starbursting EELGs at $z \simeq 1.7$
with $\log M/M_\odot \simeq 8$.
\citet{cohn:2018} carried out an analysis of SFH for EELGs at $2.5<z<4$ and
found the evidence of a recent starburst within 50~Myr,
which is also a consistent picture with other lower redshift EELGs as well as this study.
Therefore, we conclude that low-mass EELGs at $z\simeq 3.3$
are also likely to be in a starburst phase with
a short time scale of $\lesssim 100$~Myr.

Although our low-mass EELGs at $z \simeq 3.3$ show similar nature to lower redshift counterparts
in the SFR--$M_\star$ diagram and stellar ages,
one should keep in mind that
our EELG selection at a fainter magnitude of $\mathit{Ks}^\text{obs} \gtrsim 23$
may fail to select objects located between the EW cut of $\Delta m_{\mathit{Ks}}=0.3$
and the $3\sigma$ cut shown as the dash-dotted and upper solid line, respectively, in \Cref{fig:cmd}.
This means that the systematic trend of elevated SFR for low-mass EELGs at $z\simeq 3.3$ relative
to the MS could be at least partly due to a selection bias.
However, it is not straightforward to quantify the bias for \oiii emitters,
because translating \oiii flux to SFR (or Balmer line fluxes) is a complex function
of various physical parameters such as the gas-phase metallicity, ionization parameter,
hardness of UV radiation field, and electron density \citep{kewley:2013:model},
and also $\mathit{Ks}^\text{obs}$ is a combination of emission lines
and stellar continuum (i.e., stellar mass).
The \oiii/\halpha ratios derived by \citet{faisst:2016} using local analogs
show a broad distribution of $\log \oiii/\halpha \simeq -0.4$--$0.4$
which is also consistent with the range measured for SFGs at $z \simeq 2$
from the literature \citep[e.g.,][]{steidel:2014,shapley:2015:mosdef,suzuki:2016}.
On the other hand, as we will show in \autorefsec{sec:o3ew_vs_sed},
our spectroscopically identified low-mass EELGs follow
the $\text{EW} (\oiiitwo)$--$M_\star$ relation defined
for more massive normal SFGs at $z\simeq 3.3$ \citep{reddy:2018:ew}.
Therefore, it is not likely that we miss a large amount of low-mass EELGs
with significantly lower $\text{EW} (\oiiitwo)$
because of increasing photometric errors at the fainter magnitude.

\subsection{Relation between physical parameters}

\subsubsection{[\ion{O}{3}]$\lambda$5007 equivalent width and SED properties} \label{sec:o3ew_vs_sed}

\begin{figure*}[tbp]
    \centering
    \includegraphics[width=0.99\linewidth]{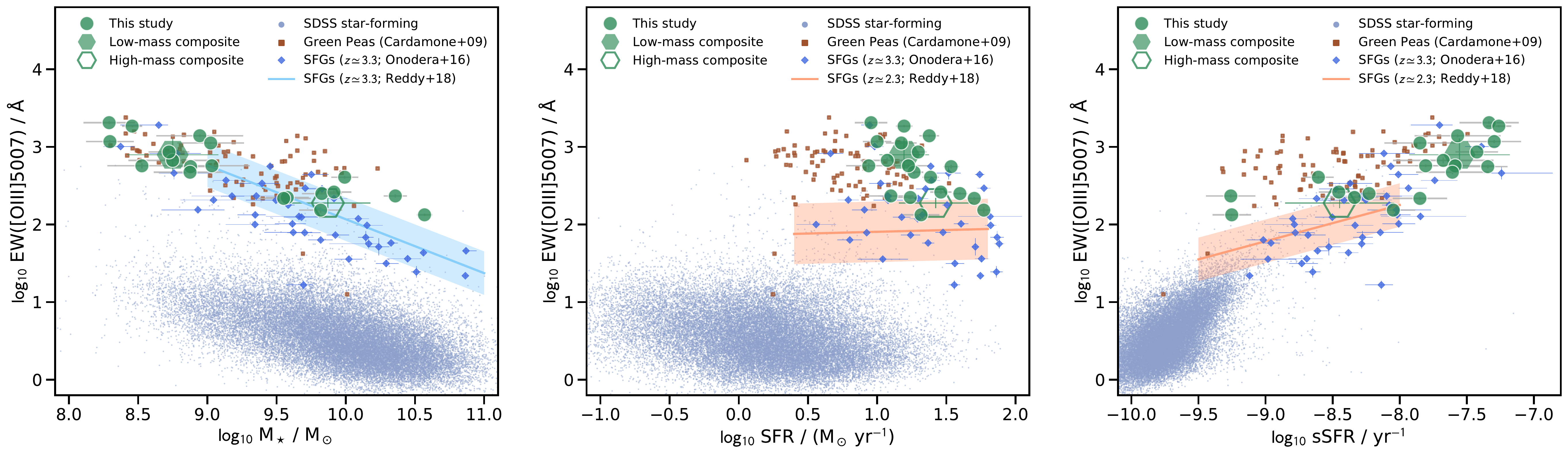}
    \caption{
        Rest-frame equivalent width of \oiiitwo as functions of
        stellar mass (\textit{left}),
        SFR (\textit{middle}),
        and sSFR (\textit{right}).
        Large green filled circles, filled hexagon, and open hexagon
        show our individual, low-mass composite, and high-mass composite EELGs at $z\simeq3.3$,
        respectively.
        Small dots correspond to local SFGs \citep{oh:2011},
        blue diamonds show normal SFGs at $z\simeq 3.3$ taken from \citet{onodera:2016},
        and filled regions show the relations derived by \citet{reddy:2018:ew}
        at $z\simeq 3.3$ in the left panel and at $z\simeq2.3$ in the middle and right panels.
        \label{fig:o3sedprop}
    }
\end{figure*}

\Cref{fig:o3sedprop} compares $\text{EW} (\oiiitwo)$ with SED properties,
namely, stellar mass, SFR, and sSFR for our EELGs at $z \simeq 3.3$
(both individual and composite measurements)
and other SFGs from the literature,
namely local SFGs selected from SDSS \citep{oh:2011},
GPs at $z\simeq 0.2$ \citep{cardamone:2009},
and normal SFGs on the MS at $z\simeq 3.3$ \citep{onodera:2016}.
We also compare them with the best-fit relationship derived by
\citet{reddy:2018:ew} for SFGs at $z\simeq 2.3$ and $3.3$.

At a fixed stellar mass, SFGs at $z\simeq 3.3$ from this study,
\citet{onodera:2016}, and \citet{reddy:2018:ew} show elevated $\text{EW} (\oiiitwo)$ by $\simeq 1.5$~dex
compared to local SFGs selected from SDSS \citep{oh:2011} as shown in the left panel of \Cref{fig:o3sedprop}.
This strong redshift evolution of the $\text{EW} (\oiiitwo)$--$M_\star$ relation
for normal SFGs
has been reported before \citep[e.g.,][]{khostovan:2016}
and the amount of the evolution in our study is similar to their result.
The linear relation for a sample of SFGs at $z\simeq 3.4$
is derived at $\log M_\star/M_\odot \gtrsim 9.5$ by \citet{reddy:2018:ew}.
They argued that the evolution
of the $\text{EW} (\oiiitwo)$--$M_\star$ relation
for the bulk of the SFG population can be explained
by the redshift evolution of the SFR--$M_\star$ and mass--metallicity relation.
Our EELG sample at $z\simeq 3.3$ with $\log M_\star / M_\odot \lesssim 9$
appears to follow the same relation extrapolated to lower stellar masses,
indicating that EELGs may be no longer a rare galaxy population
of SFGs at $z \gtrsim 3$, especially at the low stellar masses.
On the other hand, EELGs are a rare population in the local universe.
\citet{cardamone:2009} estimated the spatial number density of
GPs as 2 deg$^{-2}$ with $\mathit{r}<20.5$~mag.
Only looking at GPs, their $\text{EW} (\oiiitwo)$--$M_\star$ relation
is similar to that of EELGs and normal SFGs at $z \simeq 3.3$.

On the other hand, our EELG sample at $z \simeq 3.3$ shows up to $\simeq 3$~dex and $\simeq 1.5$~dex
higher $\text{EW} (\oiiitwo)$ at a fixed SFR compared to those of local and $z\simeq 2.3$ SFGs, respectively
(middle panel of \Cref{fig:o3sedprop}),
while the trend for GPs is quite similar to our EELGs sample
showing elevated $\text{EW} (\oiiitwo)$ at a given SFR{}.
Since \citet{reddy:2018:ew} used SFR based on \halpha luminosity,
they did not present the relationship at $z\gtrsim3$
where \halpha falls longer wavelength than the \textit{K}-band coverage.
At $z \simeq 3.3$, the MS does not seem to evolve much since $z\simeq 2.3$.
For example, the formula provided by \citet{speagle:2014} suggests
a $\simeq 0.1$~dex increase of the normalization of the MS from $z=2.3$ to $z=3.3$.
\citet{suzuki:2015} indeed found no difference in the normalization of the MS
between narrow-band selected \halpha emitters at $z=2.2$ and $2.5$ and
\oiii emitters $z=3.2$ and $3.6$, respectively.
Note that most of their narrow-band selected objects
are normal SFGs at the corresponding redshifts
as the narrow-band method selects those with
less extreme emission line strengths
than the broad-band excess method we applied.
This is also confirmed by the distribution of normal SFGs on the MS \citep{onodera:2016}
which is on average consistent with the best-fit relation at $z\simeq2.3$ by \citet{reddy:2018:ew}.
In addition, the average $\text{EW} (\oiiitwo)$ for SFGs show little evolution
at $z\simeq2$--$3$ \citep{khostovan:2016}.
Therefore, the elevated $\text{EW} (\oiiitwo)$ of our EELG sample as well as GPs can not seem to be
explained simply by the evolution of normal SFG population between $z\simeq 3.3$ to $2.3$.

Combining the above two parameters, the right panel of \Cref{fig:o3sedprop}
shows the relationship between $\text{EW} (\oiiitwo)$ and sSFR{}.
EELGs at $z\simeq 3.3$ show elevated $\text{EW} (\oiiitwo)$ at a fixed sSFR
relative to normal SFGs at $z\simeq 3.3$ \citep{onodera:2016}
and those at $z\simeq 2.3$ \citep{reddy:2018:ew},
while the slope of the relationship does not appear to
be very different between two epochs.
This trend can also be seen for GPs.
Those with lower sSFR of $\lesssim 10^{-8}\,\text{yr}^{-1}$
appear to show more elevation of $\text{EW} (\oiiitwo)$
at a given sSFR than our EELG sample.
More spectroscopic sample of EELGs at $z\simeq 3.3$ would be
required for the detailed comparison between high redshift EELGs and
local counterparts at this sSFR range.
Low-mass objects in our EELG sample with $\log M_\star/M_\odot < 9$
tend to cluster around the upper right part of the figure
as also seen from the low-mass composite point.
Normal SFGs at $z\simeq 3.3$ presented by \citet{onodera:2016} also contain objects
with elevated sSFR of $\log \text{sSFR} / \text{yr}^{-1} \gtrsim  -8$
and $\text{EW} (\oiiitwo)$ of $\gtrsim 500\,\text{\AA}$ similar to our EELGs sample.
As seen in the left panel of the figure, most of them are
low-mass objects with $\log M_\star / M_\odot \lesssim 9$.

\citet{sanders:2020:direct} studied low-mass galaxies with $\log M_\star / M_\odot \lesssim 9$
at $z=1.5$--$3.5$ with detected auroral \oiiidirect emission lines
to derive robust metallicities via the direct method.
Their sample also shows SFR above the MS at $z\simeq 2.3$ and
elevated $\text{EW} (\oiiitwo)$ similar to our EELGs at $z \simeq 3.3$.
They concluded that the auroral-line sample is younger and more metal-poor
than normal galaxies on the MS{}.
EELGs at $1.3<z<2.4$ presented by \citet{tang:2019} also occupy similar parameter space to
the auroral-line sample by \citet{sanders:2020:direct}.
Therefore, our EELGs at $z \simeq 3.3$, especially low-mass ones, are also likely to be made of galaxies
that are young and low-metallicity and
would be promising candidates for a systematic \oiiidirect follow-up spectroscopic observation.
As an example, assuming an electron density of $n_\text{e} = 500\,\text{cm}^{-3}${}
and an electron temperature of $T_\text{e} = 15000\,\text{K}$
which are reasonable values for high redshift low metallicity SFGs
\citep[e.g., ][]{shirazi:2014, kojima:2017, sanders:2020:direct},
\oiiidirect{} flux is derived as $\simeq 2$~\%{} of the \oiiitwo{} flux
by using \texttt{PyNeb}\footnote{\url{http://research.iac.es/proyecto/PyNeb/}} \citep{pyneb}.
Scaling from a typical \oiiitwo flux of our EELG sample of
$\simeq 10^{-16}\,\text{erg}\,\text{s}^{-1}\,\text{cm}^{-2}$,
their \oiiidirect{} flux is estimated to be
$\simeq 2\times 10^{-18}\,\text{erg}\,\text{s}^{-1}\,\text{cm}^{-2}$.
Our Subaru/MOIRCS setup detected the aforementioned \oiiitwo flux
with the $S/N \simeq 20$--$30$ with a $\simeq 1.5$~hr integration.
Then it will require $\gtrsim 50$--$100$ hours of telescope time
to detect \oiiidirect{} with $3\sigma$ significance from individual objects with the same instrument setup.
Therefore, the follow-up observation of \oiiidirect{} for the EELG sample
with Subaru/MOIRCS appears to be impractically expensive,
and such observations can be made more efficiently
with telescopes with larger apertures on the ground and those in space.

\subsubsection{ISM ionization properties}

\begin{figure*}
    \centering
    \includegraphics[width=0.47\linewidth]{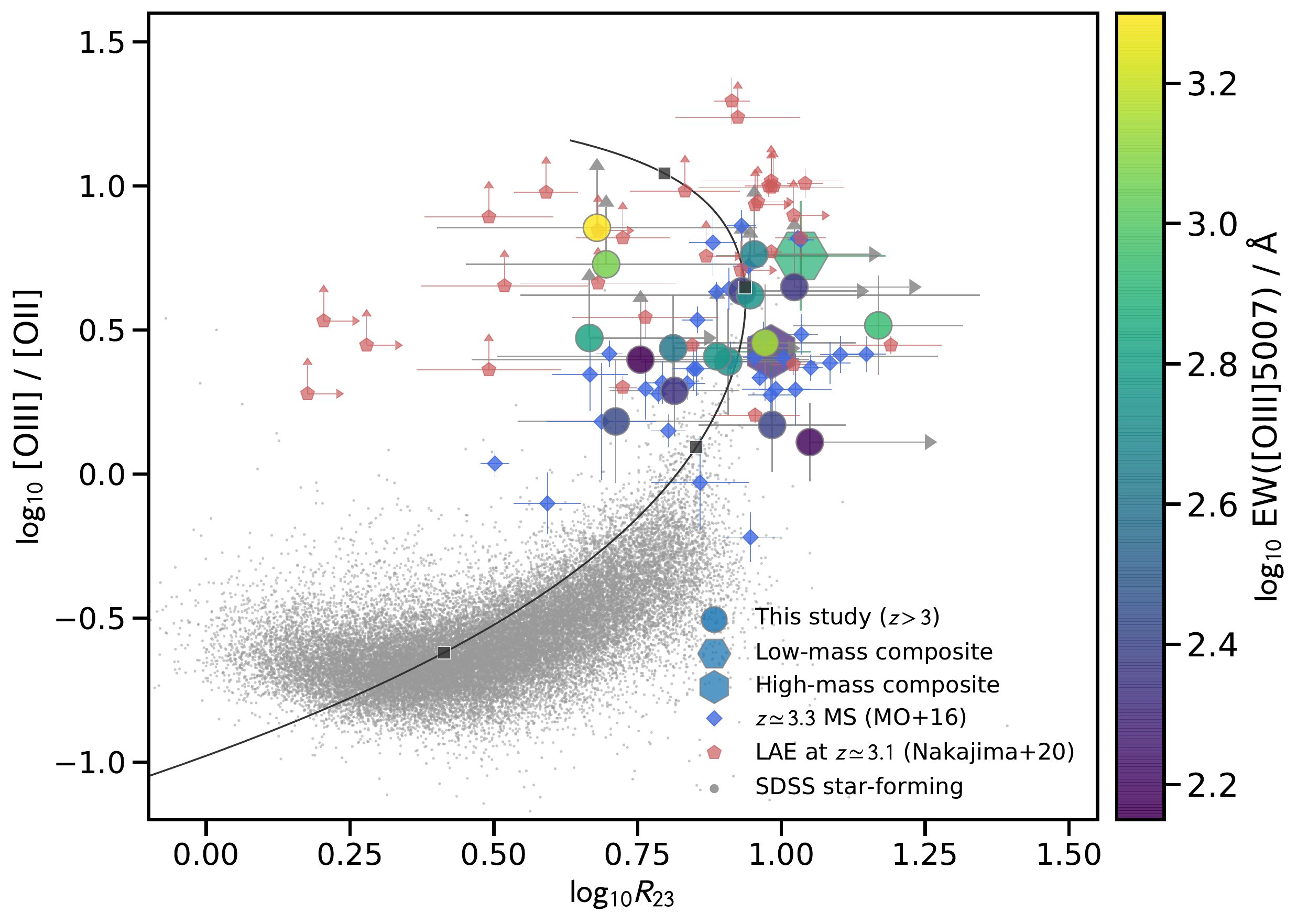}
    \includegraphics[width=0.47\linewidth]{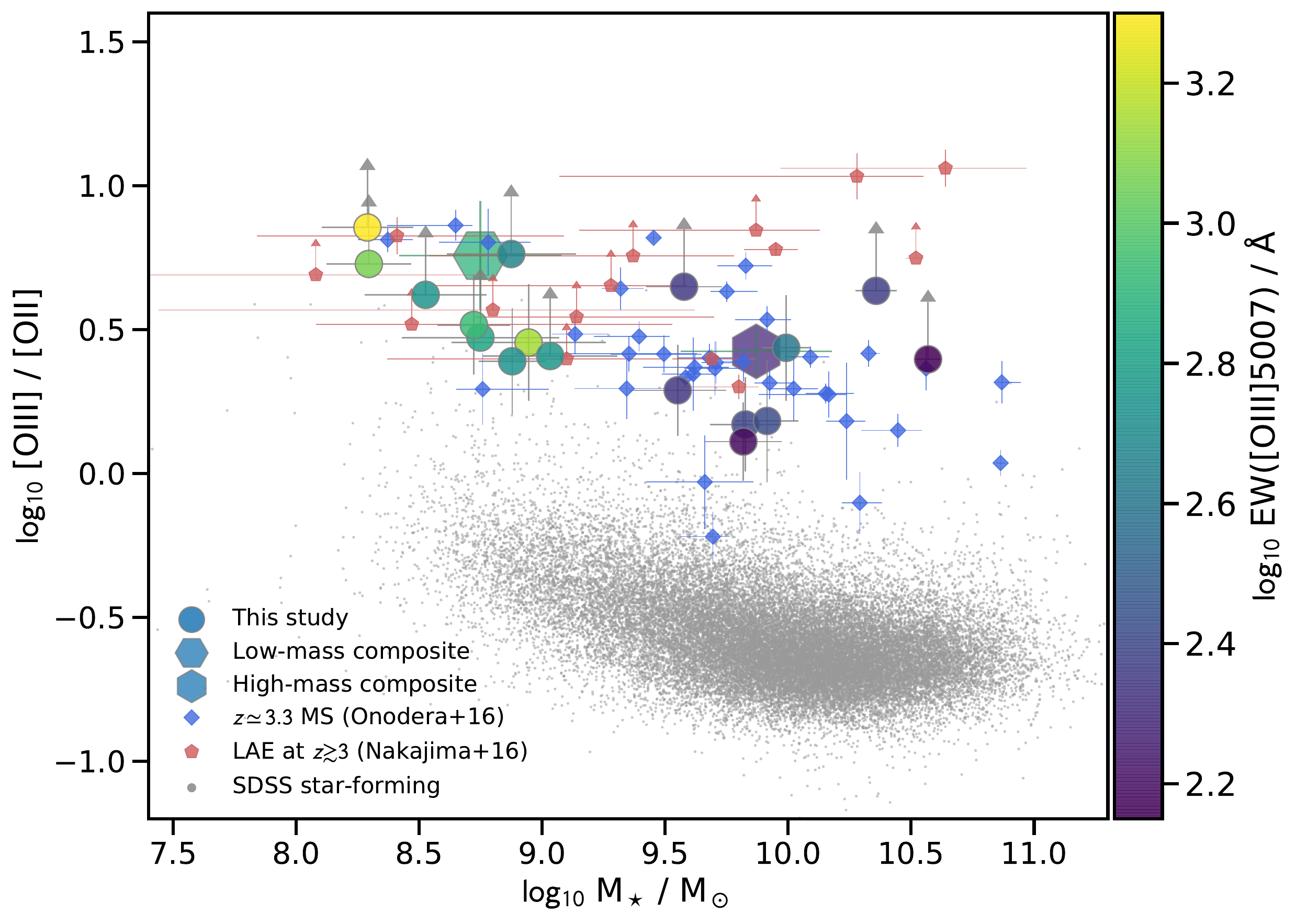}
    \caption{
        $O_{32}$ as functions of $R_{23}$ (\textit{Left}) and stellar mass (\textit{Right}).
        Our EELGs sample at $z\simeq3.3$ is shown with large circles
        color-coded by $\text{EW} (\oiiitwo)$.
        The composite values are also shown with
        a hexagon with a vertex pointing straight across
        and one with a vertex pointing directly up
        for the low-mass and high-mass composites, respectively.
        Local SFGs selected from SDSS \citep{oh:2011}
        are shown with small gray dots.
        Normal continuum-selected SFGs at $z\simeq 3.3$ \citep{onodera:2016}
        are shown with  small blue diamonds,
        while LAEs at $z \simeq 3.1$ \citep{nakajima:2020:laces}
        are shown with small red pentagons.
        In the left panel, the solid line indicates
        the relation between $O_{32}$ and $R_{23}$
        presented by \citet{maiolino:2008}
        with small black squares indicating $\ohmetal=7.5, 8.0, 8.5, \text{and}, 9.0$ from top to bottom.
        \label{fig:o32r23}
    }
\end{figure*}

\Cref{fig:o32r23} shows $O_{32}$ as functions of $R_{23}$ (left panel) and stellar mass (right panel).
$O_{32}$ is an indicator of the ionization parameter \citep[e.g.,][]{mcgaugh:1991,kewley:2002}
and $R_{23}$ is a strong-line gas-phase metallicity indicator \citep[e.g.,][]{pagel:1979,kobulnicky:2003}.
Compare to local SFGs from SDSS DR7 \citep{oh:2011},
our EELG sample at $z \simeq 3.3$ occupies the upper part of the plot,
which means that they are more metal-poor and
have higher ionization parameters.
The solid line in the left panel of \Cref{fig:o32r23} is
the locally defined relation of the two line ratios presented by \citet{maiolino:2008}
as a function of gas-phase oxygen abundance, \ohmetal.
Small squares on the line of the local relationship indicate
$\ohmetal = 7.5, 8.0, 8.5, \text{and}, 9.0$.
While the majority of local galaxies have $\ohmetal \simeq 8.5$--$9.0$,
our EELGs appear to show $R_{23}$ consistent with $\ohmetal = 7.5$--$8.5$.
EELGs at $z\simeq 3.3$ also show higher ionization parameters
compared to those of local SFGs.
The observed increase of $\simeq 1$~dex in $O_{32}$
from $z=0$ to $z\simeq 3.3$ can be translated to
an increase of $\gtrsim 1.5$~dex in the ionization parameter \citep[e.g.,][]{nakajima:2014}

In the figure, we also show normal SFGs at $z\simeq 3.3$ \citep{onodera:2016}
and LAEs at $z\simeq 3.1$ \citep{nakajima:2020:laces}.
As shown in \citet{nakajima:2014} and seen in the figure,
LAEs at $z=2$--$3$ tend to show higher $O_{32}$ indices than normal SFGs at $z\simeq 2$--$3$.
The $O_{32}$ of our entire EELGs sample at $z\simeq 3.3$
appears to be distributed between the two populations.

It is also noticeable that EELGs with higher $O_{32}$ values
are more likely to show larger $\text{EW} (\oiiitwo)$.
As shown in \Cref{fig:o3sedprop}, there is an anti-correlation between $\text{EW} (\oiiitwo)$ and galaxy stellar mass.
Massive EELGs at $z\simeq 3.3$ with $\log M_\star / M_\odot \gtrsim 9.5$
typically have $\text{EW}(\oiiitwo) \lesssim 300$~\AA{},
while lower mass ones show $\text{EW}(\oiiitwo) \gtrsim 500$~\AA{}.
From the right panel of \Cref{fig:o32r23},
it is clearer that at $z\gtrsim 3$ the distribution of the two line ratios considered here
for massive EELGs are similar to normal SFGs
and that for low-mass ones are closer to LAEs \citep[see also][]{suzuki:2017}.
These trends are also seen for the composite measurements.

\subsubsection{Ionizing photon production efficiency} \label{sec:xiion}

\begin{figure*}
    \centering
    \includegraphics[width=0.9\linewidth]{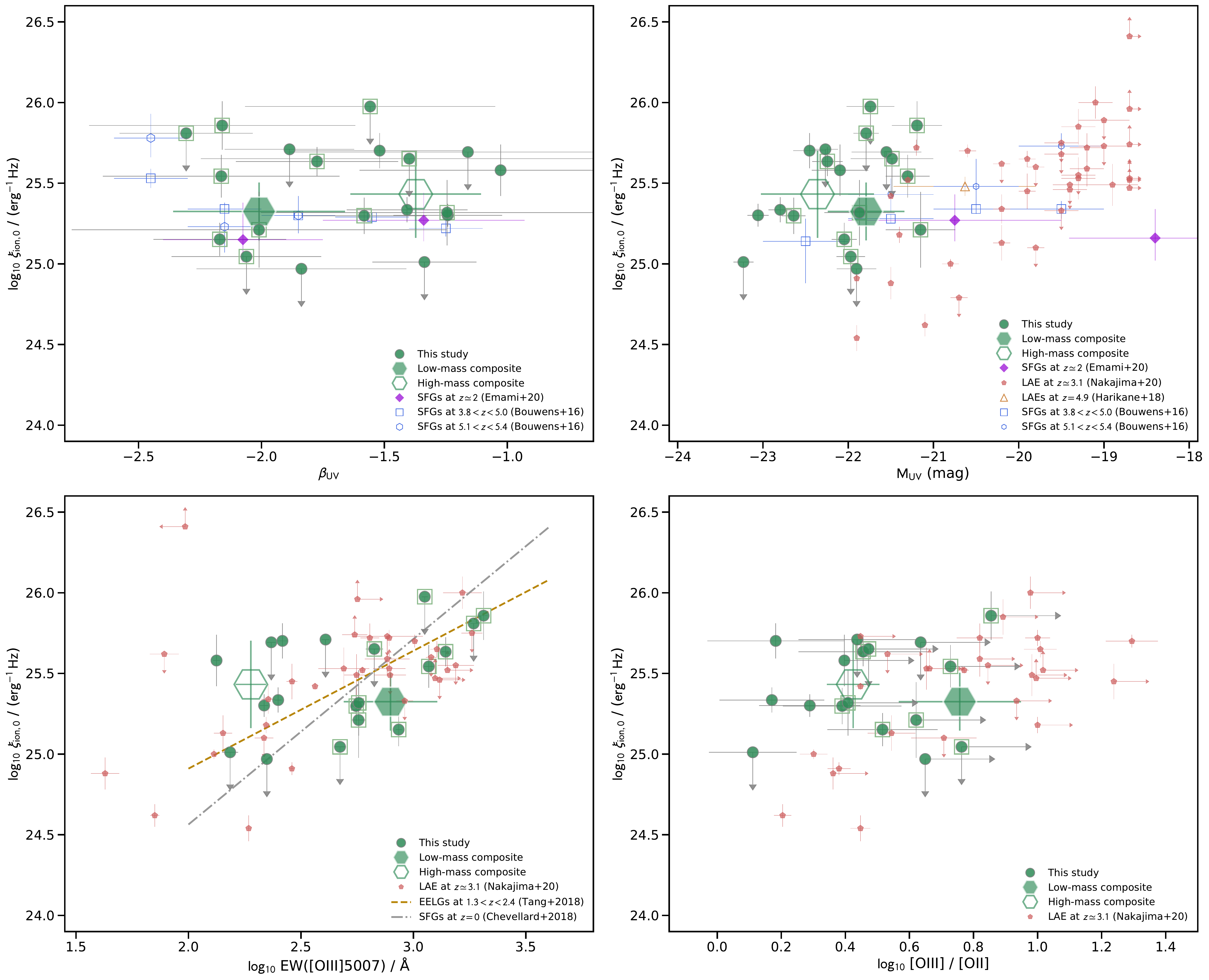}
    \caption{
        The ionizing photon production efficiency \xiionz as
        a function of
        the UV spectral slope $\beta_\text{UV}$ (\textit{Top Left}),
        the absolute UV magnitude at 1500~\AA{} $M_\mathrm{UV}$ (\textit{Top Right}),
        $\text{EW} (\oiiitwo)$ (\textit{Bottom Left}),
        and $O_{32}$ (\textit{Bottom Right}).
        Our EELGs at $z \simeq 3.3$ are shown with green circles
        and those with $\log M_\star / M_\odot \lesssim 9$ are
        highlighted with squares.
        Filled and open hexagons represent the low-mass and high-mass composites
        of the sample, respectively.
        Filled diamonds show stacked measurements of
        low-mass SFGs at $1.4<z<2.7$ \citep{emami:2020}.
        We adopt their \textit{Standard} stacking,
        and  plot the mid points of each bin along the horizontal axes
        taken from their Table 1.
        LAEs at $z=3.1$ taken from \citet{nakajima:2020:laces} are shown
        with small red pentagons.
        Measurements for color-selected SFGs at $3.8 < z < 5.0$ and $5.1 < z < 5.4$
        from \citet{bouwens:2016} are shown with open blue squares and hexagons, respectively.
        A stacked measurement for LAEs at $z=4.9$ are shown with an open triangle \citep{harikane:2018}.
        The relationship derived for EELGs at $1.3 < z < 2.4$ \citep{tang:2019} is
        shown with the dashed line.
        When provided, we consistently used values for the SMC curve for dust correction.
        \label{fig:xiion}
    }
\end{figure*}

In the upper left panel of \Cref{fig:xiion},
we compare \xiionz as a function of \uvbeta.
Our EELG sample at $z \simeq 3.3$ shows a flat relation
between the two parameters with a median value of
$\log \xiionz / (\xiionunit)= 25.54$,
though 8 out of 19 objects have upper limits for \xiionz.
This is also the case for the composite spectra. \citet{emami:2020} have found a similar trend,
i.e., no \uvbeta dependence of \xiionz
for a sample of low-mass ($M_\star=10^{7.8}$--$10^{9.8}\,M_\odot$) galaxies at $z \simeq 2$
\citep[see also][for a sample of \halpha emitters]{matthee:2017}.
They corrected dust attenuation by using the SMC curve for the UV luminosity from SED fitting
and the Milky Way (MW) extinction curve \citep{cardelli:1989}
for the \halpha luminosity based on the Balmer decrement.
On the other hand, different trends have also been reported in the literature.
\citet{bouwens:2016} reported a $\simeq 0.3$~dex increase of
mean \xiionz for the bluest galaxies ($\uvbeta < -2.3$)
for a sample of sub-$L^{*}$ SFGs at $z=4$--$5$,
while a constant value of $\log \xiionz / (\xiionunit)= 25.34$
was measured at $\uvbeta > -2.3$.
Their conclusion was not affected by the choice of
the dust attenuation curve (Calzetti or SMC).
\citet{shivaei:2018} found a similar trend to \citet{bouwens:2016} for a sample of
SFGs at $z \sim 2$ from the MOSDEF survey \citep{kriek:2015:mosdef}
using the Calzetti curve for dust correction.
When using the SMC curve, their relationship becomes
almost flat with a minimum of \xiionz at $\uvbeta \simeq -1.5$ \citep{shivaei:2018}.

To check whether our result changes by using the Calzetti curve
instead of the SMC curve, we derived \xiionz assuming
the Calzetti curve for stellar continuum following the
conversion from \uvbeta to \ebvstar
presented in \citet{reddy:2018},
and the MW extinction curve
for nebular emission lines with a conversion
${E(\mathit{B}-\mathit{V})}_{\text{neb,MW}} = 1.34 {E(\mathit{B}-\mathit{V})}_\text{star,Calz}$
derived by \citet{theios:2019}.
Similar to \citet{shivaei:2018},
\xiionz becomes $0.1$--$0.4$~dex smaller with the Calzetti curve for dust correction
than with the SMC curve with a median difference of $0.28$~dex.
The difference appears to become larger for higher \uvbeta objects
as also reported by \citet{shivaei:2018},
but there seems still no correlation between \xiionz and \uvbeta,
likely to be partly due to the small number statistics with
a large number of upper limits and a relatively large scatter of the distribution of the sample.
In addition to the choice of the attenuation curves for dust correction,
the conversion factor between the nebular to stellar \ebv
could be functions of galaxy properties such as stellar mass, SFR, and hence sSFR
as seen in the local universe \citep{koyama:2019}.
Although it is not clear whether such relation can be applicable to
galaxies at $z \simeq 3.3$ with intense \oiii emission lines,
this would potentially introduce an additional uncertainty in \xiionz.

The main reason why our EELGs at $z \simeq 3.3$ show
no dependence of \xiionz on \uvbeta
could be their selection as the most intense $\oiii+\hbeta$ emitting objects.
Unlike our selection method, other studies mentioned above did not impose
any cuts in emission line strengths.
In fact, objects with $\log \xiionz / (\xiionunit) \simeq 26$
exist at all \uvbeta up to $\uvbeta \simeq -1$
in the tail of the distribution in \citet{shivaei:2018}.

\smallskip

The \xiionz of our EELG sample as well as
those of SFGs at $z\simeq 2$ \citep{emami:2020},
LAEs at $z\simeq 3.1$ \citep{nakajima:2020:laces},
SFGs at $z\simeq4$--$5$ \citep{bouwens:2016},
and LAEs at $z=4.9$ \citep{harikane:2018}
are shown as a function of UV absolute magnitude $M_\text{UV}$
in the upper right panel of \Cref{fig:xiion}.
LAEs at $z=4.9$ from \citet{harikane:2018} follow
the trend of the $\xiionz$--$M_\text{UV}$ relation
of SFGs at $z \simeq 4$--$5$ \citep[see also \citealt{lam:2019}]{bouwens:2016}.
Our EELGs at $z\simeq 3.3$ show relatively bright $M_\text{UV}$
similar to brighter objects of \citet{bouwens:2016},
but have on average higher \xiionz by up to $\simeq 0.8$~dex.
This may be also an indication of the extreme nature of our sample
based on the selection with significant \textit{K}-band excesses.

Comparing to LAEs at $z\simeq 3.1$ from \citet{nakajima:2020:laces},
our EELGs show a similar range of \xiionz,
but brighter UV luminosities by $2$--$3$~mag
despite the similar stellar mass range.
A part of the large difference in $M_\text{UV}$
could be due to the assumption of
a uniform $\ebvstar = 0.01$
based on the SED fitting by \citet{fletcher:2019}.
Note, however, that even a small increase of the dust attenuation
parameter can result in a significant increase of the UV luminosity
especially when using the SMC extinction curve.
If we assume that our EELGs are extinction free,
$M_\text{UV}$ becomes fainter on average by $\simeq 1.1$~mag.
Therefore, the difference in dust correction methods
cannot fully explain the difference in UV luminosities
between two samples and the reason why the LAE selection can pick up
fainter objects with similar stellar masses.
Another possible reason of the difference is that \citet{fletcher:2019} performed
the SED fitting by using BPASS v2.1 stellar population models \citep{eldridge:2017},
while we adopted BC03 models in the \cigale framework.
\citet{eldridge:2017} pointed out that BC03 models are redder in UV at very young ages of $< 10$~Myr
and also redder in red-optical to near-IR wavelength at ages more than a few times $100$~Myr.
The choice of different population synthesis models can result in a systematic effect
in estimating stellar masses.
SFGs at $z\simeq2$ by \citet{emami:2020} include objects
with much fainter UV luminosities, $M_\text{UV}\gtrsim -18$
and they found no correlation between \xiionz and $M_\text{UV}$
in contrast to \citet{nakajima:2020:laces}.
Note that to derive the UV luminosities
\citet{emami:2020} used BC03 templates and the SMC curve for dust correction
which are different from those used in \citet{nakajima:2020:laces},
and that \citeauthor{emami:2020} sample
shows on average smaller $\text{EW} (\oiiitwo)$ of $\simeq 100$~\AA{}
than that of \citet{nakajima:2020:laces} of $600$--$1000$~\AA{},
which may be reasons of the discrepancy at the faintest UV magnitude bin.

\smallskip

A strong correlation between \xiionz and $\text{EW} (\oiii)$ has been claimed
for high redshift EELGs \citep{emami:2020, tang:2019,nakajima:2020:laces}
as well as nearby SFGs \citep{chevallard:2018}.
The lower left panel of \Cref{fig:xiion} shows
the relation between \xiionz and $\text{EW} (\oiiitwo)$
for our EELG sample at $z \simeq 3.3$ along with that
for LAEs at $z\simeq 3.1$ \citep{nakajima:2020:laces}
and the best-fit relations derived for EELGs at $1.3 < z < 2.4$ \citep{tang:2019}
and for local EELGs \citep{chevallard:2018}.
Our EELG sample at $z \simeq 3.3$ follows the similar
trend with the relationships in the literature.
The correlation between \xiionz and $\text{EW} (\oiii)$
appears to be the strongest among four parameters shown in \Cref{fig:xiion}.
On the other hand, the composite data do not show such correlation.
This is probably because we adopted the inverse-variance weighting,
which results in putting more weights on objects with
the strongest, i.e., the highest S/N, emission line fluxes.
As shown in the figure, our EELGs with $\text{EW(\oiiitwo)} \gtrsim 300\,\text{\AA}$
are exclusively low-mass galaxies with $\log M_\star / M_\odot \lesssim 9$,
which can also be clearly seen in the composite spectra.
Moreover, those with $\text{EW(\oiiitwo)} \gtrsim 1000\,\text{\AA}$
show particularly high \xiionz of $\gtrsim 25.5$,
suggesting that those with low stellar mass and high $\text{EW} (\oiii)$ are
very efficient in producing LyC photons.

\smallskip

In the lower right panel of \Cref{fig:xiion},
we show the relationship between \xiionz and $O_{32}$.
EELGs at $z\simeq3.3$ do not show an apparent correlation
unlike LAEs at $z \simeq 3.1$ \citep{nakajima:2020:laces}
as shown with small red pentagon symbols.
Quantitatively, the Spearman's rank correlation coefficients are 0.05 and 0.48
and the corresponding two-sided \textit{p}-values are 0.84 and 0.001,
for our EELGs at $z\simeq 3.3$ and LAEs at $z\simeq 3.1$, respectively.
\citet{shivaei:2018} also found a 0.3~dex enhancement of \xiion
at large $\log O_{32} \gtrsim 0.5$, while \xiion is constant at $\log O_{32} \lesssim 0.5$.
They also found that this enhancement can be observed also at low $\log O_{32}$ galaxies
when the SMC extinction curve is used for dust correction instead of the Calzetti curve.
The systematic effect due to dust correction could be one of the reasons
why there is no correlation between \xiionz and $O_{32}$ for our EELG sample.
Another reason would be again the biased selection of EELGs as the strongest \oiii emitters.
Compared to LAEs at $z \simeq 3.1$, our EELGs show similar \xiionz, but systematically
lower $O_{32}$ values by $\sim 0.2$~dex,
though a relatively large number of objects with only limits
hampers a detailed comparison of the distribution of two populations.
Because $O_{32}$ can be used as indicators of ionization parameters
and gas-phase metallicity \citep[e.g.,][]{maiolino:2008},
it is expected that objects with larger $O_{32}$
are efficient hydrogen ionizing photon producers and show elevated \xiion values.
As seen before, most of the EELGs with larger $\log O_{32}$ of $\gtrsim 0.4$ are
low-mass galaxies with $\log M_\star/M_\odot\lesssim 9$
which follow the distribution of LAEs at a similar redshift well.

\smallskip

In summary, our EELG sample at $z \simeq 3.3$ are efficient hydrogen ionizing photon producers,
in particular at lower stellar masses of $\log M_\star / M_\odot \lesssim 9$.
The relationships between \xiionz and various observed properties (\Cref{fig:xiion})
suggest that our EELGs have similar properties to luminous SFGs at $z\gtrsim 4$
\citep{bouwens:2016} and EELGs at $1.4 \lesssim z \lesssim 2.5$ \citep{tang:2019}.
Comparison with LAEs at $z\simeq 3.1$ indicates that LAEs are systematically
less luminous in UV, and more highly ionized objects than the EELGs at $z \simeq 3.3$.
Among the investigated parameters,
\xiionz of our EELGs show a marginal correlation with $\text{EW} (\oiii)$,
while other parameters does not impact on the \xiionz.
This would be mainly due to our biased selection of objects with large $\text{EW}(\oiii+\hbeta)$.
Another possible source of systematics is the choice of the attenuation curve for dust correction.
If we use the Calzetti curve instead of the SMC curve,
\xiionz are reduced by $0.1$--$0.4$~dex.
Although the change is more prominent in objects with larger \uvbeta,
the conclusions of the comparison above are still unchanged.
The fact that many of our EELG objects do not
have detected \hbeta and \oii also dilute correlations if any.
A deeper spectroscopic follow-up observation \citep[e.g.,][]{nakajima:2020:laces}
is required to make a more detailed comparison.

\subsection{Implications for the Lyman continuum escape}

\begin{figure}[tbp]
    \centering
    \includegraphics[width=0.98\linewidth]{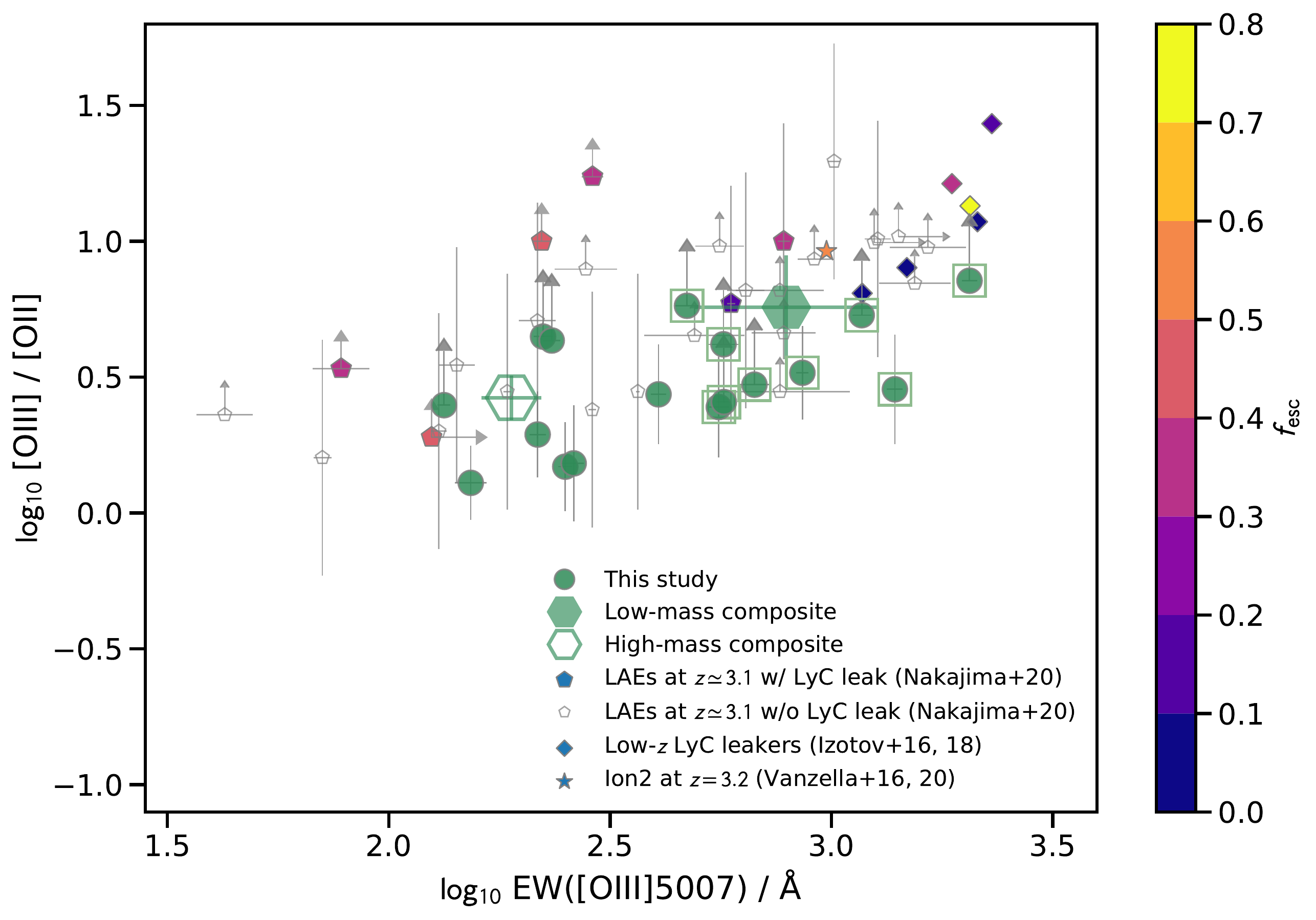}
    \caption{
        $O_{32}$ as a function of $\text{EW} (\oiiitwo)$.
        Our EELGs at $z\simeq 3.3$ are shown as filled green circles
        of which low mass ($\log M_\star / M_\odot \lesssim 9$) ones
        are indicated by open squares.
        Stacked measurements are shown with open and filled hexagons
        for low-mass and high-mass composites, respectively.
        Filled pentagons, diamonds, a small square, and star symbol
        are previously confirmed from the literature color-coded by LyC escape fraction (\fesc);
        LAEs at $z\simeq 3.1$ \citep{nakajima:2020:laces},%
        low redshift galaxies at $z \simeq 0.3$ \citep{izotov:2016:nat,izotov:2018:o32},
        and a high redshift LyC leaker at $z=3.2$
        \citep{vanzella:2016:fesc50,vanzella:2020:starclusters}, respectively.
        Open gray pentagons show LAEs at $z\simeq 3.1$ with no LyC detection \citep{nakajima:2020:laces}.
        \label{fig:o32o3ew_lyc}
    }
\end{figure}

Our primary motivation to search for EELGs with intense \oiii emission at $z>3$ is
to investigate a population of galaxies that can be considered as analogs which contributed to
reionize the universe in the EoR as close as possible to the epoch,
but not too high redshift where the universe is fully opaque to the LyC emission.
Indeed, we have shown that we were able to select such galaxies at $z\simeq 3.3$
with $\text{EW} (\oiiitwo)$ reaching up to $\simeq 2000\,\text{\AA}$ efficiently by the \textit{Ks}-band excess method.
Such large $\text{EW}(\hbeta+\oiii)$ values are comparable to those inferred for SFGs
at the EoR
\citep[e.g.,][]{smit:2014,smit:2015,robertsborsani:2016,stark:2017,debarros:2019,endsley:2020}.

In addition to such large $\text{EW} (\oiii)$, we have shown that the EELG sample at $z\simeq 3.3$ in this study
are characterized with enhanced SFR (and sSFR) compared to the MS,
large $O_{32}$ values, low level of dust attenuation,
bright in UV, and high \xiionz,
suggesting that they are young, low metallicity galaxies in a bursty phase of star formation
with highly ionized ISM{}.
These physical properties are more pronounced in the less massive ones,
similar in many aspects to LAEs at a similar redshift and blue SFGs at even higher redshifts
as well as EELGs at lower redshifts.
Some of these galaxies in the literature are found to show various degrees of LyC escape.
Here, we compare observational properties which we have in hand in this study
and are considered as indicators of the LyC leakage.

\Cref{fig:o32o3ew_lyc} shows the relation between $O_{32}$ and $\text{EW} (\oiiitwo)$ for our EELG sample at $z\simeq 3.3$
and confirmed LyC leakers at $z\simeq 0.3$ \citep{izotov:2016:nat, izotov:2018:o32},
and \textit{Ion2} at $z=3.2$ \citep{vanzella:2016:fesc50,vanzella:2020:starclusters}.
We also overplot LAEs at $z=3.1$ presented by \citet{nakajima:2020:laces}
including those without LyC detection.
LyC leakers are color-coded by the LyC escape fraction (\fesc).
It is notable that there is a large diversity in \fesc among the LyC leakers
from a few percent to $>70$~\%{},
and they tend to have large $O_{32}$ values of $\gtrsim 3$--$5$.

Compared to LyC leakers shown in \Cref{fig:o32o3ew_lyc},
our EELGs follow the similar $O_{32}$--$\text{EW}(\oiiitwo)$ relationship to confirmed LyC leakers,
though many of our EELGs have only lower limits in $O_{32}$ (i.e., non-detection in \oii).
Our EELGs at $z\simeq 3.3$ show large $O_{32}$ values of $\gtrsim 3$ for about a half of the sample
and most of them are low-mass galaxies with $\log M_\star / M_\odot \lesssim 9$
as also seen in the composite measurements.
Also, as shown before, these low-mass EELGs show large $\text{EW}(\oiiitwo) \gtrsim 500$~\AA{}.

The comparison makes it tempting to classify these EELGs at $z \simeq 3.3$ as LyC leakers with large values of \fesc.
Indeed, photoionization models suggest that $O_{32}$ can be used as an indicator of \fesc \citep[e.g.,][]{jaskot:2013,nakajima:2014},
and observational data also appear to follow the predictions \citep[][and references therein]{faisst:2016}.
However, recent studies suggest that having large $O_{32}$ values is a necessary condition, not a sufficient condition,
possibly due to geometrical effects along the line-of-sight
\citep{izotov:2018:o32,naidu:2018,jaskot:2019,nakajima:2020:laces}.
\citet{bassett:2019} have made a detailed comparison
of the the relation between \fesc and $O_{32}$
with density-bounded models using the MAPPINGS~V photoionization code
\citep{allen:2008:mappings3,sutherland:2018:mappings5}.
They found that $O_{32}$ shows the largest variation at $\fesc < 0.1$,
while it seems to provide little constraint at larger \fesc than $\simeq 0.1$,
especially for low-metallicity cases.
The fact that many of LAEs without LyC detections have comparably large
$O_{32}$ to LyC leakers also supports these scenarios \citep{nakajima:2020:laces}.
Yet, EELGs with very large $O_{32}$ of $\gtrsim 10$ are highly likely to contain LyC leakers,
because such high $O_{32}$ ratios can be emerged from a density-bounded nebula
(i.e., fully ionized gas cloud)
from which LyC emission can escape easily compared to an ionization-bounded nebula
where the size of the nebula is determined
by the {Str{\"{o}}mgren} sphere \citep{nakajima:2014,jaskot:2019}.

\begin{figure}[tbp]
    \centering
    \includegraphics[width=\linewidth]{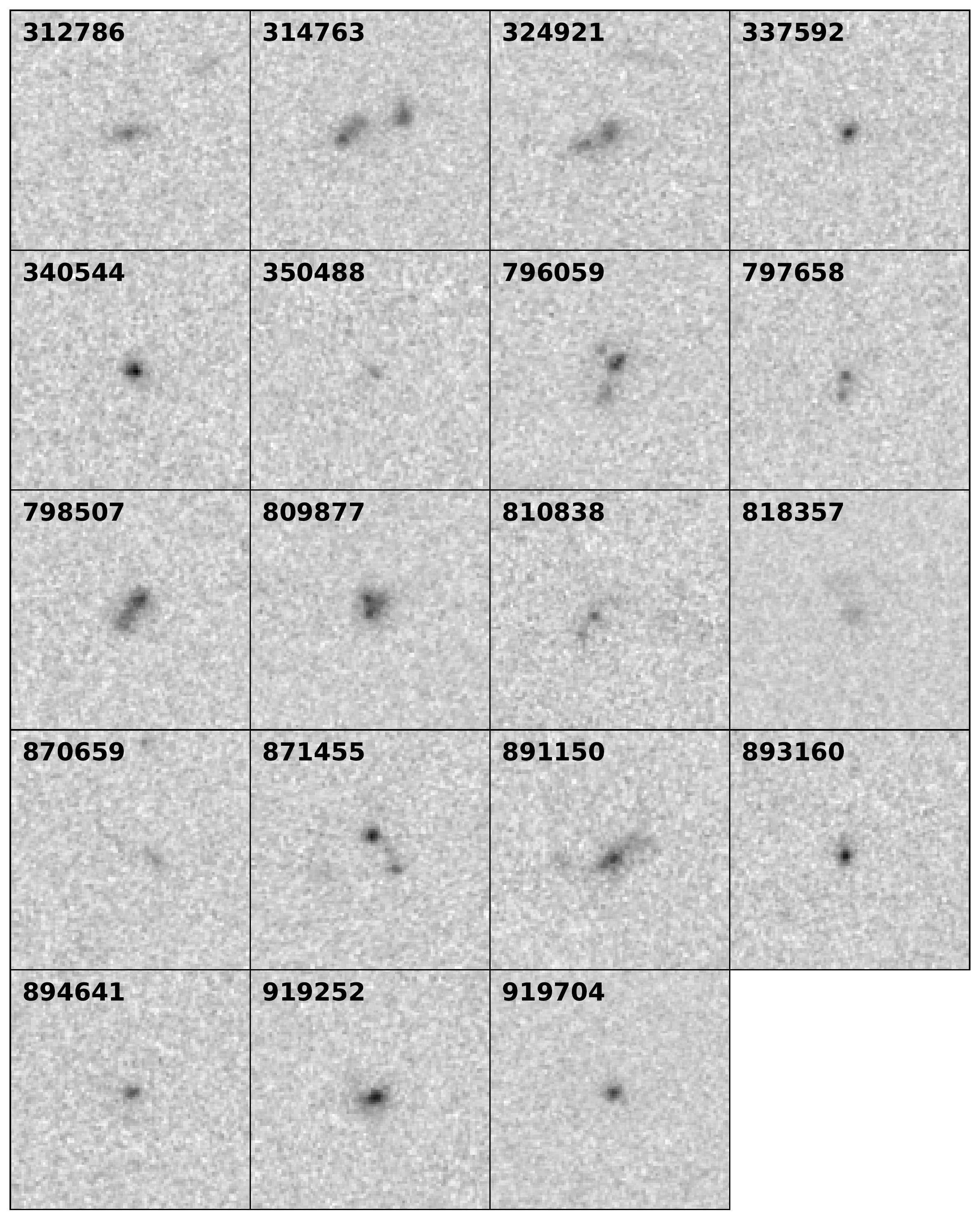}
    \caption{
        Cutouts of \textit{HST}/ACS F814W mosaic \citep{koekemoer:2007,massey:2010}
        for the spectroscopically confirmed EELGs at $z\simeq 3.3$.
        Center of each image corresponds to the coordinate provided in
        the COSMOS2015 catalog (i.e., measured on the UltraVISTA image),
        and the size of the image is $3'\times 3'$.
        The image is scaled by the arcsinh stretch \citep{lupton:1999,lupton:2004}.
        \label{fig:acsimg}
    }
\end{figure}

Additionally, \citet{bassett:2019} have also discussed that
gas-rich galaxy mergers may play a role to break the $O_{32}$--\fesc relation
by re-distributing gas in the system.
Such mergers of low-mass metal-poor galaxies or inflow of metal-poor gas into the system
are suggested to trigger bursty star formation by a sudden change in the gas accretion rate,
which could result in breaking the equilibrium of self-regulation of star formation
\citep{lilly:2013,onodera:2016,amorin:2017}.
Alternatively, other mechanisms that are not necessarily classified as mergers
but minor interactions and violent disk instabilities,
can also produce star-forming clumps
and cause mixing of gas in galaxies \citep{ribeiro:2017}.
Such clumpy structures are often observed in
low-mass metal-poor EELGs at $z<1$ \citep[e.g.,][]{kunth:1988,lagos:2014, amorin:2015, calabro:2017}.
Among 11 EELGs in our sample with $\log M_\star / M_\odot \lesssim 9$,
five of them show indications of on-going merger or clumpy structure,
while the rest shows compact, point-like morphology in \textit{HST}/ACS F814W images
shown in \Cref{fig:acsimg}.
The apparent diversity of morphology of our EELG sample at $z \simeq 3.3$
could partly reflect different merger stages.
Note, however, that these merger- or clumpy-like structures can be
due to chance projections along the line-of-sight.
In order to fully characterize the nature of individual components of each object,
deep high resolution multiband imaging and integral field spectroscopy
using currently available facilities such as \textit{HST} and ground-based 8--10m class telescopes
with adaptive optics, and future ones such as \textit{James Webb Space Telescope} (\textit{JWST})
and ground-based 30m-class telescopes will be required.

\subsection{Implications for galaxies in the epoch of reionization}\label{sec:eor}

Whether it is possible for star-forming galaxies in the EoR to
provide sufficient hydrogen ionizing photons (i.e., LyC photons)
escaped from them is crucial to assess the dominant ionizing
population of the reionization.
Important parameters here are the hydrogen ionizing photon production efficiency in the case of no LyC escape \xiionz
and LyC escape fraction \fesc as the product of the two parameters determines the amount of
LyC photons outside of the system.
As a canonical value,  $\log \xiionz / (\xiionunit) = 25.2$--$25.3$ has been widely used \citep[e.g.,][]{robertson:2013}.
The value is suggested from BC03 stellar population models at about the solar metallicity
with a young stellar population with $\simeq 1$--$10$~Myr old ages for a single burst star formation history.
It is also observationally supported
by the analysis of \uvbeta slope of galaxies at $z\simeq 7$--$9$
\citep{dunlop:2013,robertson:2013}.

As seen in \Cref{fig:xiion} and \Cref{tab:ismprop},
the majority of our EELGs at $z \simeq 3.3$ have $\log \xiionz  / (\xiionunit)$
larger than the canonical value, though they include those with upper limits.
This means that the majority of our EELGs are more efficiently producing LyC photons
than expected by the BC03 models with the aforementioned parameters.
Therefore, in order to explain the elevated values of \xiionz,
different properties of massive stars in galaxies would be needed,
such as the IMF, SFH, inclusion of binary populations, and metallicity.
\citet{shivaei:2018} have made a detailed investigation
on the effect of these parameters using the BPASS and BC03 models.
Their fiducial model consists of the Salpeter IMF \citep{salpeter:1955} with
the lower and upper mass cut of $0.5\,M_\odot$ and $100\,M_\odot$, respectively,
and the slope $\alpha=-2.35$,
constant SFH over 300~Myr, $Z=0.15\,Z_\odot$, and without binary populations,
resulting in $\log \xiion / (\xiionunit) = 25.23$.
In their analysis, BC03 models with the same parameterization as the fiducial model
give $\log \xiion / (\xiionunit) = 25.06$,
$\sim 0.1$~dex lower than the canonical value mentioned above.
Including the binary population, increasing the upper mass cut of IMF to $300\,M_\odot$,
increasing the IMF slope to $\alpha=-2.00$, and lowering the metallicity to $0.07\,Z_\odot$
result in the increase of the \xiion
by $\sim 0.17$~dex, $\sim 0.16$~dex, $\sim 0.18$~dex, and $0.02$~dex, respectively
\citep[see Table 1 in][]{shivaei:2018}.
Since our EELGs at $z\simeq3.3$ typically have $\log \xiionz / (\xiionunit) \simeq 25.5$
and as large as $\simeq 26.0$, a combination of more than one changes from the fiducial model
would be required to explain these values.

In addition to the effects related to their stellar populations as discussed above,
the non-zero LyC escape fraction can scale the \xiionz by $1 / (1 - \fesc)$.
Statistically speaking, $\fesc=10$--$20$~\%{} is supposed to be necessary
for galaxies to fully ionize the universe at $z>6$ \citep[e.g.,][]{ouchi:2009,robertson:2013},
and in this case the intrinsic \xiion can be even larger by $\simeq 0.05$--$0.1$~dex.
If we simply use the best-fit relation between $O_{32}$ and \fesc derived by \citet{faisst:2016:lyc},
$\fesc=0.05, 0.1$, and $0.45$ correspond to $O_{32}=3, 5$, and $10$, respectively,
which can be translated to the increase of
the intrinsic \xiion of 0.02~dex, 0.05~dex, and 0.26~dex, respectively.
Therefore, even a moderate escape of LyC from our EELGs can further increase the discrepancy.
For those with large values of $O_{32} \simeq 10$ and \xiionz,
the intrinsic \xiion can be as large as $\log \xiion / (\xiionunit) \gtrsim 26$,
which cannot be explained even by the most extreme set of parameters for BPASS models considered by \citet{shivaei:2018}.
As suggested by \Cref{fig:sfrmass} and \Cref{fig:o32r23},
even younger stellar population age with a burst-like star formation history, lower metallicity,
or a combination of them would be required to explain the most extreme objects in our sample.
Note that, as discussed in the previous section, the LyC escape does not seem to be determined
solely by the ionization properties inferred by for example $O_{32}$ and $\text{EW} (\oiii)$,
but the geometry of ionized gas is suggested to be crucial \citep[e.g., ][]{nakajima:2020:laces}.

\section{Summary} \label{sec:summary}

We presented results from a systematic search for EELGs at $z \simeq 3.3$.
The selection was done by using the \textit{Ks}-band excess flux
due to intense $\hbeta + \oiiitot$ emission lines relative to the best-fit
stellar continuum models.
The \textit{Ks}-excess method was verified successfully by the subsequent
near-IR spectroscopic follow-up with Subaru/MOIRCS after the detector upgrade.
We observed 23 of 240 \textit{Ks}-excess objects and
identified 21 objects among which 19 galaxies are confirmed to be at $3<z<3.6$
with rest-frame $\text{EW}(\oiiitwo) > 100\,\text{\AA}$ and up to $\simeq 2000\,\text{\AA}$.
Focusing on the spectroscopically identified EELGs at $z \simeq 3.3$,
we investigated their physical properties
derived based both on rich multiband photometry in the COSMOS field
and on the MOIRCS near-IR spectra.
Our main results can be summarized as follows.

\begin{itemize}
    \item Photometric redshifts for EELGs at $z \simeq 3.3$ listed in the widely used COSMOS2015 catalog
          agree with the MOIRCS spectroscopic redshifts,
          despite the insufficient treatment of nebular emission lines in the catalog.
          However, stellar masses from the catalog are overestimated for low-mass objects
          compared to those derived with our dedicated SED fitting including
          more realistic treatment of the emission line contribution.
    \item AGN do not seem to dominate the EELG population at $z\simeq 3.3$ judged based on
          their mid- to far-IR SEDs, emission line widths, and \oiii/\hbeta emission line ratios.
    \item Our spectroscopically confirmed EELGs at $z\simeq 3.3$ are
          on the MS at stellar masses of $\gtrsim 10^{9.5}\,M_\odot$,
          while the lower-mass EELGs show elevated SFR than the MS by $\gtrsim 0.5$--$1.0$~dex.
          This suggests that low-mass EELGs are in a bursty star formation phase
          with young stellar population age of $\lesssim 100$~Myr.
    \item EELGs at $z\simeq3.3$ appear to follow the $M_\star$--$\text{EW}(\oiiitwo)$ relation
          of normal SFGs at the similar redshift, while at fixed SFR and sSFR
          they show higher $\text{EW}(\oiiitwo)$ than normal SFGs at $z\simeq 2.3$ and $3.3$,
          also supporting the idea that they are young and low metallicity.
    \item A strong-line metallicity indicator $R_{23}$ indicates that
          gas-phase metallicity of EELGs at $z\simeq 3.3$ are $\ohmetal = 7.5$--$8.5$,
          while an ionization parameter sensitive index $O_{32}$
          shows on average $\sim 1$~dex increase relative to local SFGs,
          translated to $\gtrsim 1.5$~dex increase of the ionization parameter.
          Comparison to other galaxy populations at $z\gtrsim 3$,
          our EELGs show similar ionization properties
          to normal SFGs at high masses ($\gtrsim 10^{9.5}\,M_\odot$)
          and LAEs at low masses ($\lesssim 10^{9}\,M_\odot$).
    \item The hydrogen ionizing photon production efficiency, \xiionz, shows
          a positive correlation with $\text{EW} (\oiiitot)$.
          On the other hand, no significant correlation with \xiionz is found for
          the UV spectral slope, UV luminosity, and $O_{32}$,
          possibly reflecting the extreme nature of our EELGs compared to objects from the literature
          with different selections such as LAEs and dropouts.
    \item Comparison of the $O_{32}$--$\text{EW(\oiiitwo)}$ relation
          between our EELG sample and SFGs with measured LyC leakage
          suggests that our EELGs with the largest $O_{32}$ values of $\gtrsim 10$
          and $\text{EW(\oiiitwo)}\gtrsim 1000\,\text{\AA}$ likely caused
          by density-bounded nebulae can be the most promising
          LyC leaker candidates.
    \item Our EELGs show \xiionz larger than the commonly used canonical value,
          suggesting that they are efficient in producing ionizing photons to ionize
          their surrounding ISM{}.
          Considering that \xiionz of our EELG sample at $z\simeq 3.3$
          is similar to that of SFGs at higher redshifts of $z>4$--$5$,
          galaxies at the EoR are likely to have such large values of \xiionz
          as inferred from the broad-band observations.
          Assuming an average LyC escape fraction of $\sim 10$~\%{},
          intrinsic \xiion can become $\log \xiion/(\xiionunit) \gtrsim 26$.
          In order to reproduce the elevated \xiion values,
          different properties of massive stars in galaxies than the canonical models
          would be required,
          including top-heavy IMF, burst-like SFH with younger age,
          inclusion of binary populations, lower metallicity, or
          a combination of them.
\end{itemize}


\acknowledgments

We are grateful to the anonymous referee for providing constructive comments
which improved the manuscript significantly.
We thank Nao Fukagawa to help the observation,
Emanuele Daddi, Andreas Faisst, Taysun Kimm, and Hyewon Suh for fruitful discussions,
and the staff of Subaru Telescope for supporting the observations.
This work was supported by JSPS KAKENHI Grant-in-Aid for Young Scientists (B) Grant Number JP17K14257.
This research made use of Astropy,  
a community-developed core Python package for Astronomy \citep{astropy:2013, astropy:2018}
and APLpy, an open-source plotting package for Python \citep{aplpy:2012,aplpy:2019}.
The authors wish to recognize and acknowledge the very significant cultural role and reverence that the summit of Maunakea has always had within the indigenous Hawaiian community.  We are most fortunate to have the opportunity to conduct observations from this mountain.
%

%

\vspace{5mm}
\facilities{Subaru (MOIRCS)}


\software{
    APLpy \citep{aplpy:2012, aplpy:2019},
    Astropy \citep{astropy:2013,astropy:2018},
    CIGALE \citep{burgarella:2005, noll:2009, boquien:2019},
    EAZY \citep{brammer:2008},
    emcee \citep{emcee},
    IRAF \citep{tody:1986:iraf,tody:1993:iraf},
    lmfit \citep{lmfit},
    matplotlib \citep{matplotlib},
    MCSMDP \citep{yoshikawa:2010},
    Numpy \citep{numpy:2020},
    PyNeb \citep{pyneb},
    seaborn \citep{seaborn-latest},
    TOPCAT \citep{topcat}
}

\end{document}